\documentclass[lettersize,journal]{IEEEtran}
\usepackage{amsmath,amsfonts}
\usepackage{algorithmic}
\usepackage{algorithm}
\usepackage{array}
\usepackage[caption=false,font=normalsize,labelfont=sf,textfont=sf]{subfig}
\usepackage{textcomp}
\usepackage{stfloats}
\usepackage{url}
\usepackage{verbatim}
\usepackage{graphicx}
\usepackage{cite}
\usepackage{tikz}
\usepackage{multirow}
\usepackage{booktabs}
\usepackage{bigstrut}
\usepackage{hyperref}

\begin{document}

\title{Over-The-Air Federated Learning: Status Quo, Open Challenges, and Future Directions}

\author{Bingnan Xiao, Xichen Yu, Wei Ni,~\IEEEmembership{Senior Member,~IEEE}, Xin Wang,~\IEEEmembership{Fellow,~IEEE}, \\and H. Vincent Poor,~\IEEEmembership{Life Fellow,~IEEE}

\thanks{B. Xiao, X. Yu, and X. Wang are with the Key Laboratory for Information Science of Electromagnetic Waves (MoE), Department of Communication Science and Engineering, Fudan University, Shanghai 200433, China. E-mail: xwang11@fudan.edu.cn.

W. Ni is with the Data61, Commonwealth Scientific and Industrial Research Organisation (CSIRO), Sydney, NSW 2122, Australia. Email: wei.ni@data61.csiro.au.

H. Vincent Poor is with the Department of Electrical and Computer
Engineering, Princeton University, Princeton, NJ 08544, USA (email: poor@princeton.edu).
}}



\maketitle

\begin{abstract}
The development of applications based on artificial intelligence and implemented over wireless networks is increasingly rapidly and is expected to grow dramatically in the future. The resulting demand for the aggregation of large amounts of data has caused serious communication bottlenecks in wireless networks and particularly at the network edge. Over-the-air federated learning (OTA-FL), leveraging the superposition feature of multi-access channels (MACs), enables users at the network edge to share spectrum resources and achieves efficient and low-latency global model aggregation. This paper provides a holistic review of progress in OTA-FL and points to potential future research directions. Specifically, we classify OTA-FL from the perspective of system settings, including single-antenna OTA-FL, multi-antenna OTA-FL, and OTA-FL with the aid of the emerging reconfigurable intelligent surface (RIS) technology, and the contributions of existing works in these areas are summarized. Moreover, we discuss the trust, security and privacy aspects of OTA-FL, and highlight concerns arising from security and privacy. Finally, challenges and potential research directions are discussed to promote the future development of OTA-FL in terms of improving system performance, reliability, and trustworthiness. Specifical challenges to be addressed include model distortion under channel fading, the ineffective OTA aggregation of local models trained on substantially unbalanced data, and the limited accessibility and verifiability of individual local models.

\end{abstract}


\section{Introduction}

\IEEEPARstart{T}he envisioned sixth-generation (6G) of mobile communication systems has attracted significant attention in academic and industrial communities~\cite{10002946}. An important trend in discussions of 6G is a shifting of machine learning (ML) tasks from central cloud infrastructures to the network edge, capitalizing on the computational potential of edge devices and the flexibility of network connectivity \cite{peltonen20206g}, \cite{eldar2022machine}. Federated learning (FL), a distributed learning framework, is particularly well-suited for edge applications~\cite{10118940,10000870,9920736,9779339}. Initially proposed in \cite{mcmahan2017communication}, FL has recently gained considerable traction. In FL settings, geographically distributed users train their own models using local data and then transmit their local model parameters or gradients to a base station (BS) for model aggregation. The BS subsequently returns the obtained global model to the users, repeating this process until model convergence~\cite{10123399}. Unlike traditional centralized learning settings, FL does not necessitate the transmission of large amounts of training data, thereby reducing communication costs and helping ensure data privacy to a significant extent~\cite{9880724}.

\subsection{Over-The-Air Federated Learning}

The concept of over-the-air (OTA) computation, also known as AirComp, was introduced in \cite{4305404} to leverage the signal superposition characteristics of wireless multiple access channels (MACs) for function computation. OTA computation offers the advantage of resource consumption reduction since the BS only needs to handle functions uploaded by users rather than individual data. Dedicated radio resource allocation for each user is unnecessary, making the OTA communication-computation approach highly suitable for the model aggregation process in FL.

\begin{figure}[!h]
	\centering
	\includegraphics[width=1\linewidth]{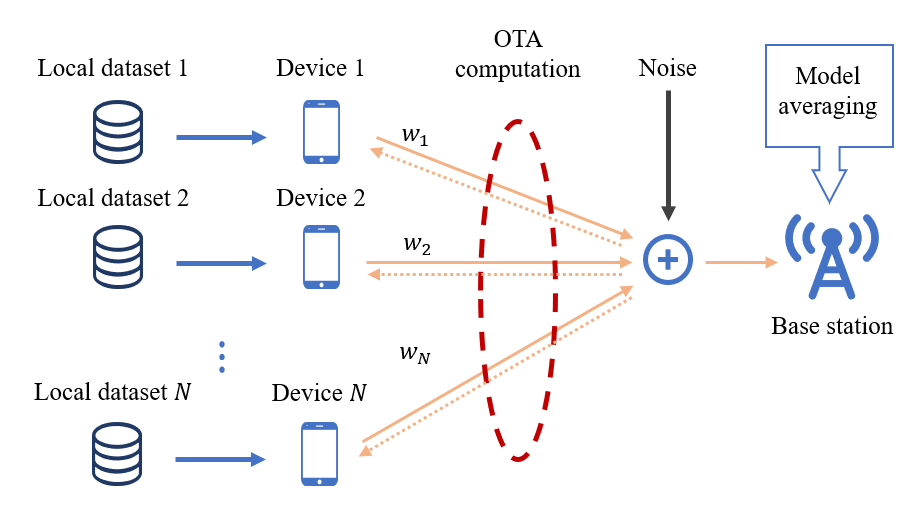}
	\caption{An illustration of the system model of OTA-FL. Users perform several local updates and then transmit their respective models updates through MAC. Upon the OTA aggregation, the updates are sent to the BS for the global model update. }
	\label{fig:The system model of OTA-FL}
\end{figure}

In recent years, a growing body of research has explored the signal superposition capability of OTA for aggregating local models transmitted by users in the context of wireless federated learning, commonly referred to as OTA-FL. OTA-FL enables users to share the same spectral resources, enhancing communication efficiency. In OTA-FL, edge devices can simultaneously transmit their local model updates, aggregating the models over the air in a ``one-time" manner, as illustrated in Fig. \ref{fig:The system model of OTA-FL}. However, due to the inherent effects of channel fading and additive noise, the aggregated signal received at the BS inevitably exhibits some bias. Consequently, it becomes necessary to devise suitable transmission and reception strategies to mitigate the impact of channel fading and noise, thereby improving the convergence of OTA-FL systems.

As an emerging technique, the consideration of trust, security, and privacy in OTA-FL has been limited despite their paramount importance for the sustainability and reliability of ML models, as unveiled in this survey. While those aspects have been extensively studied regarding the traditional FL \cite{9415623}, it is not straightforward to apply or extend the existing solutions to OTA-FL due to its distinct communication and aggregation strategy. For example, individual local models are obsolete in OTA-FL. As a consequence, the methods for detecting adversarial local models, e.g., Krum or multi-Krum, would no longer be applicable. 

\subsection{Contribution and Organization}
    This survey provides a detailed and comprehensive survey of existing studies on OTA-FL, addresses significant challenges, and outlines potential future research directions. The key contributions of the survey are listed as follows.

    \begin{itemize}
    \item 
    We provide a systematic classification of OTA-FL from a fresh perspective, including single-antenna OTA-FL, multi-antenna OTA-FL, and OTA-FL with the assistance of reconfigurable intelligent surfaces (RISs).
    We holistically summarize the optimization objectives and algorithms used in the existing works, highlighting their commonalities and pros and cons.
    
    \item
    We delineate the trust, security, and privacy risks confronted by OTA-FL, uncover the gap in the existing literature and point to novel research perspectives crucial for user concerns.
    
    \item
    We identify critical challenges and future directions that need to be addressed, including efficient aggregation, stringent synchronization requirements, abundant data heterogeneity, trustworthiness, and communication-learning metrics.
    \end{itemize}
It is found through this survey that OTA-FL is still in its infancy. Existing studies have typically aimed to minimize the optimality gap of OTA-FL for individual model aggregations under the assumption of ML models with (strongly) convex and smooth loss functions. Little consideration has been given to understanding the impact of channel fading processes on OTA-FL, especially under multi-antenna settings. Moreover, there has been little consideration of the practical implementation of OTA-FL. Challenges, such as model distortion under channel fading, the ineffective OTA aggregation of local models trained on substantially unbalanced data, and the limited accessibility and verifiability of individual local models, have yet to be addressed. 
    
    The rest of this paper is structured as follows. Section II categorizes and discusses single-input single-output (SISO) OTA-FL system designs from the perspectives of user selection and power control. Section III extends the SISO architecture to multiple antenna settings and summarizes strategies addressing optimization problems under multi-antenna settings. In Section IV, the role of RISs in OTA-FL is discussed, and the corresponding system design strategies are analyzed. Section V describes the trust, security and privacy issues and their impacts on OTA-FL. Section VI points out the challenges and future research directions of OTA-FL, followed by conclusions in Section VII. The abbreviations involved in the paper are collated in Table I.

    \textit{Notation}: Boldface upper- and lower-cases stand for matrix and vector, respectively; $\mathbb{R}^N$ denotes the space of all $N\times1$ real-valued column vectors; $\mathbb{E}[\cdot]$ takes mathematical expectation; $h^{E}$ denotes the effective channel gain.
   
\begin{table}
\label{table0}
\caption{List of abbreviations}
% \resizebox{\textwidth}{!}{
\begin{tabular*}{\linewidth}{ll}
\toprule
\textbf{Abbreviation} & \textbf{Full form}                        \\ \midrule
6G                    & The sixth generation                      \\
AI                    & Artificial Intelligence                   \\
BEV                   & Best Effort Voting                        \\
BS                    & Base Station                              \\
CSI                   & Channel State Information                 \\
DAB                   & Decomposed Aggregation Beamformer         \\
DC                    & Difference-of-convex Function             \\
DNN                   & Deep Neural Network                       \\
DP                    & Differential Privacy                      \\
DRIS                  & Double-Reconfigurable Intelligent Surface \\
FL                    & Federated Learning                        \\
GAN                   & Generative Adversarial Network            \\
I.I.D.                & Independent Identically Distributed       \\
ISAC                  & Integrated Sensing and Communication      \\
LDP                   & Local Differential Privacy                \\
mmWave                & Millimeter Wave                           \\
MAC                   & Multiple Access Channel                   \\
MIMO                  & Multiple-input Multiple-output            \\
ML                    & Machine Learning                          \\
MSE                   & Mean Square Error                         \\
MTPDD                 & Mixed-timescale penalty-dual-decomposition\\
NIP                   & Non-linear Integer Programming            \\
NOMA                  & Non-Orthogonal Multiple Access                \\
NN                    & Neural Network                            \\
Non-I.I.D.            & Non-Independent Identically Distributed   \\ 
OMA                   & Orthogonal Multiple Access                \\
OMA-FL                & Orthogonal multiple access Federated Learning \\
OTA                   & Over-the-air                              \\
OTA-FL                & Over-the-air Federated Learning           \\
RIS                   & Reconfigurable Intelligent Surface        \\
SCA                   & Successive Convex Approximation           \\
SDR                   & Semidefinite Relaxation                   \\
SGD                   & Stochastic Gradient Descent               \\
SISO                  & Single-input Single-output                \\
SNR                   & Signal-to-noise Ratio                     \\
S-CSI                 & Statistical Channel State Information     \\
WPT                   & Wireless Power Transfer                   \\ \bottomrule
\end{tabular*}
% }
\end{table}

\section{SISO OTA-FL}

We start with an overview of OTA. The primary aim of OTA is to integrate the concurrently transmitted local models or gradients for computing a specific set of nomographic functions \cite{6557530}:
\begin{equation}\label{Nomographic Function}
F\left(d_1, d_2, \ldots, d_K\right)=\psi\left(\sum_{k=1}^K \varphi_k\left(d_k\right)\right),
\end{equation}
where $F(\cdot): \mathbb{R}^K\rightarrow\mathbb{R}$ is a nomographic function, $d_k\in\mathbb{R}$ is a data sample at user $k$; $\varphi_k(\cdot): \mathbb{R} \rightarrow \mathbb{R}$ and $\psi(\cdot): \mathbb{R} \rightarrow \mathbb{R}$  denote the pre-processing function and the post-processing function, respectively. From \eqref{Nomographic Function}, we can observe that the nomographic function represents the process of signal transmission and aggregation in OTA-FL, i.e., each user's local data $d_k$ undergoes pre-processing through $\varphi_k(\cdot)$ and is transmitted over wireless channels. The post-processing is subsequently performed through $\psi(\cdot)$ at the BS.

Currently, studies on OTA-FL are still in their infancy, with the majority of studies focused on SISO OTA-FL. Critical issues, such as user selection and power control, have been the primary concerns. There are also studies dedicated to the joint design of user selection and power control to address multiple goods at the same time.
In Table \ref{siso table}, we categorize the existing studies on SISO OTA-FL into three classes, including user selection, power control and joint design, and summarize their design goals and optimization strategies, as well as their strengths and weakness. 

\subsection{User Selection}

In SISO OTA-FL systems, several factors, such as the size of local data and channel quality, often influence the significance of local updates for each user. In scenarios where aggregating data from all users is not feasible, selecting the most ``important" users for training participation becomes necessary, as this selection can significantly impact the system's performance. However, the user selection problem is typically a discrete non-convex problem, which necessitates the search for efficient solving methods~\cite{9796935}. 

A trade-off is proposed in \cite{8870236}, where the authors suggest that aggregating training data from a more significant number of devices per round can expedite convergence. However, this approach may also lead to increased aggregation error due to the inclusion of devices with poorer channels. 
The work presented in \cite{9605599} addresses a user scheduling problem considering the cumulative energy budget of each user over $T$ rounds. By analyzing the convergence performance, the authors design an estimated drift-plus-penalty algorithm using Lyapunov optimization. An estimation method is employed to forecast the norm of local gradients to overcome the challenge of unknown communication energy.

In the context of multiple parallel OTA-FL in cellular networks, discussed in \cite{9844173}, the authors focus on a scenario where a server handles multiple model training tasks from distinct groups utilizing identical radio resources. They define an optimization problem to simultaneously optimize receiver combiner vectors and user selection to minimize the time-varying convergence upper bound. To tackle this problem, they decompose it into two sub-problems, which are addressed using a successive convex approximation (SCA) method and a greedy algorithm. 

The authors of \cite{10038617} propose a dynamic device scheduling framework using channel inversion-based power control. They design a measurement factor inspired by the convergence upper bound, which takes into account both the quality and quantity of selected users as the optimization objective. The problem is then solved using Lyapunov optimization.

\subsection{Power Control}
The design of power control strategies plays a crucial role in mitigating the impact of fading and ensuring robust received signal strength for users in weak channels at the BS, thereby enhancing the performance of OTA-FL systems. Existing research primarily focuses on power control issues through the minimization of the optimality gap and obtaining the optimal power distribution factor. Most existing works, such as \cite{cao2022transmission}, \cite{9606731}, and \cite{9843892}, aim to achieve user power control by minimizing the optimality gap in each round. 

In \cite{cao2022transmission}, an optimality gap minimization method is proposed to address aggregation errors characterized by mean square error (MSE). This is accomplished by optimizing transmission power and a denoising factor, and the problem is solved using an alternating minimization algorithm. Similarly, in \cite{9606731}, a power control strategy is obtained by minimizing the optimality gap while considering unbiased aggregation constraints. The formulation incorporates both average and maximum power constraints for each user, and convex reformulations are employed with structured optimal solutions. In \cite{9843892}, the authors derive the effects of model aggregation errors accumulated among communication rounds using the upper limit of the time-average norm of model parameter gradients. Power control and transceiver policies are obtained by minimizing the derived upper bound through an alternating optimization algorithm. 
In \cite{9791337}, OTA-FL is investigated in a multi-cell wireless network where different training tasks are performed in each cell. A convergence analysis is carried out by taking inter-cell interference, and a problem is formulated to minimize the error gap across all cells while adhering to power constraints.

In order to decrease the overhead of channel estimations in OTA-FL, the authors of \cite{9780892} put forward two OTA-FL schemes that utilize statistical channel state information (CSI), known as S-CSI. These schemes aim to reduce the associated cost of channel estimations.
\nocite{10039388}
The convergence bound of OTA-FL has been analyzed in \cite{9076343}, with a specific focus on the impact of channel fading on the distortion of OTA-FL. A few interesting insights are unveiled:
\begin{itemize}
    \item The per-round convergence upper bound depend primarily on the mean and variance of channel fading states, $\mu_h$ and $\sigma_h^2$.

    \item With the increase of $\mu_h$ and the decrease of $\sigma_h^2$, this convergence upper bound can effectively shrink when the ML model has strongly convex or convex loss functions. Meanwhile, a larger $\mu_h$ and a smaller $\sigma_h^2$ can counteract the side effects caused by channel noise.
\end{itemize}
In light of the insights, an optimal adaptive power control scheme has been recently proposed to combat the channel-induced model distortion of OTA-FL by leveraging S-CSI in~\cite{Our-Paper}. 
The objective of this scheme is to minimize the optimal gap of OTA-FL under any unknown channel fading conditions, as given by
\begin{equation}\begin{aligned}
    &\mathop {\min }\limits_{\{ \rho(h)\geq0,\forall h\} } G\left( \mu_{h^E},\sigma_{h^E}^2 \right) &\text{s.t.}\hspace{0.2cm} \mathbb{E}_h\left[ \rho^2(h) \right]\leq P_0 ,
\end{aligned}\end{equation}
where $G(\cdot)$ is the optimality gap of the convergence, and $\rho(h)$ specifies the power control policy. 
Based on the findings in~\cite{9076343}, this optimization problem can be translated to essentially construct an efficient channel with a large mean $\mu_{h^E}$ and a small variance $\sigma_{h^E}^2$.
By transforming the problem into a nested optimization problem and solving it with the Lagrange dual method, the optimal power control policy can be found to exhibit the following structure: 
\begin{equation}\label{lemma-3}
    \rho^*(h)=\frac{h(2\mu^*_{h^{E}}-\nu^*)}{2(h^2+\lambda^*)},\hspace{0.3cm}\forall h.
\end{equation}  
where the optimal dual variables $\lambda^*$ and $\nu^*$ is obtained using a subgradient ascent method, and $\mu_{h^{E}}^*$ is obtained efficiently using a one-dimensional search method.

Following this policy, at each communication round, a user only needs to know its current channel state and update the dual variables accordingly. Even without the knowledge of future channel variations, the policy has been proven to provide a long-term optimal power control strategy under independent and identically distributed (I.I.D.) channel environments.  
The policy can also be extended in the situation where even the channel statistics, i.e., $\mu_{h^E}$ and $\sigma_{h^E}^2$, are unknown \textit{a-priori}. The policy can estimate the channel statistics on-the-fly based on the accumulated historical observations of the channels.

\subsection{Joint Design of User Selection and Power Control}
There has been a growing interest in simultaneously considering and optimizing user selection and power control for OTA-FL, recognizing that joint design may not always guarantee optimality for each objective due to transformations or decompositions of the original problems. However, this joint approach can enhance system robustness by improving the system design from multiple perspectives.

In \cite{fan2021joint} and \cite{guo2022joint}, an optimal co-design strategy for user selection and power control is obtained by minimizing the optimality gap, leveraging system convergence analysis. This joint strategy takes into account both user selection and power control to achieve improved performance. Furthermore, in~\cite{10001136}, the authors extend the consideration to include energy harvesting strategies on the user side. They formulate a non-convex non-linear integer programming (NIP) problem to optimize client selection on a per-training-round basis, addressing the energy harvesting aspect of OTA-FL in addition to user selection and power control optimization.

\begin{table*}[hbt] 
    \centering
    \renewcommand\arraystretch{1.2}    
    \caption{Summary of the existing studies on SISO OTA-FL systems}
    \label{siso table}
    \resizebox{\textwidth}{!}{
\begin{tabular}{|c|c|c|c|c|} 
\hline
\textbf{Category}                                                             & \textbf{Objective function}                                                                             & \textbf{Method}                                                                                                                                 & \textbf{Pros}                                                                                                                                           & \textbf{Cons}                                                                                                                                                   \\ 
\hline
\multirow{3}{*}{\begin{tabular}[c]{@{}c@{}}User \\Selection\end{tabular}}     & \multirow{2}{*}{\begin{tabular}[c]{@{}c@{}}Minimize \\optimality gap\end{tabular}}                      & \begin{tabular}[c]{@{}c@{}}Lyapunov optimization \\framework \cite{9605599}\end{tabular}                                                        & \begin{tabular}[c]{@{}c@{}}The framework considers \\a long-term \\energy-constrained scenario.\end{tabular}                                            & \begin{tabular}[c]{@{}c@{}}The heuristic objective function\\~makes it difficult to \\obtain the optimal solution.\end{tabular}                                 \\ 
\cline{3-5}
                                                                              &                                                                                                         & \begin{tabular}[c]{@{}c@{}}Alternating minimization \\algorithm and greedy algorithm \\after problem transformation~\cite{9844173}\end{tabular} & \begin{tabular}[c]{@{}c@{}}Multiple groups of users \\are within the system, \\each training one ML task.\end{tabular}                                  & \begin{tabular}[c]{@{}c@{}}Both of the subproblems \\have suboptimal solutions\\after problem transformation.\textcolor[rgb]{0.216,0.255,0.318}{}\end{tabular}  \\ 
\cline{2-5}
                                                                              & \begin{tabular}[c]{@{}c@{}}Maximize  the quantity \\ and qualities of devices\end{tabular}                   & \begin{tabular}[c]{@{}c@{}}Lyapunov optimization \\framework~\cite{10038617}\end{tabular}                                                       & \begin{tabular}[c]{@{}c@{}}The scheme designs\\~a residual feedback mechanism\\~to fully utilize local updates.\end{tabular}                            & \begin{tabular}[c]{@{}c@{}}Some variables need to \\be determined~in advance,~\\making it difficult to \\detect their optimal values.\end{tabular}              \\ 
\hline
\multirow{5}{*}{\begin{tabular}[c]{@{}c@{}}Power \\Control\end{tabular}}      & \multirow{3}{*}{\begin{tabular}[c]{@{}c@{}}Minimize \\optimality gap\end{tabular}}                      & \begin{tabular}[c]{@{}c@{}}Alternating minimization \\algorithm after \\problem transformation~\cite{cao2022transmission}\end{tabular}          & \begin{tabular}[c]{@{}c@{}}It quantifies the latency \\advantage of~OTA-FL\\compared to OMA-FL.\end{tabular}                                            & \begin{tabular}[c]{@{}c@{}}The conclusions drawn \\need to be based on \\strongly convex cases.\end{tabular}                                                    \\ 
\cline{3-5}
                                                                              &                                                                                                         & \begin{tabular}[c]{@{}c@{}}Lagrange dual method \\ after convexification~\cite{9606731}\end{tabular}                                            & \begin{tabular}[c]{@{}c@{}}~It is able to obtain \\the optimal solution\\by convexifying the problem.\end{tabular}                                      & \begin{tabular}[c]{@{}c@{}}The scheme assumes that\\~the user end can only \\perform one local update.\end{tabular}                                             \\ 
\cline{3-5}
                                                                              &                                                                                                         & \begin{tabular}[c]{@{}c@{}}Lagrange dual method \\ and subgradient algorithm~\cite{Our-Paper}\end{tabular}                                      & \begin{tabular}[c]{@{}c@{}}The algorithm effectively \\utilizes S-CSI and \\can be executed online.\end{tabular}                                        & \begin{tabular}[c]{@{}c@{}}It requires the assumption \\that~fading channels \\for each user are~I.I.D.\end{tabular}                                            \\ 
\cline{2-5}
                                                                              & \begin{tabular}[c]{@{}c@{}}Minimize \\time-average MSE\\ (derived from \\optimality gap)\end{tabular}   & \begin{tabular}[c]{@{}c@{}}Alternating minimization \\algorithm~\cite{9843892}\end{tabular}                                                     & \begin{tabular}[c]{@{}c@{}}A DNN solver is designed \\to simplify the process of\\algorithmic solving.\textcolor[rgb]{0.216,0.255,0.318}{}\end{tabular} & \begin{tabular}[c]{@{}c@{}}The interpretability of \\the DNN solver needs \\to be further elucidated.\end{tabular}                                              \\ 
\cline{2-5}
                                                                              & \begin{tabular}[c]{@{}c@{}}Minimize \\error induced gaps\\ (derived from \\optimality gap)\end{tabular} & \begin{tabular}[c]{@{}c@{}}Feasibility check \\ after problem resolution~\cite{9791337}\end{tabular}                                            & \begin{tabular}[c]{@{}c@{}}It provides~\\performance analysis \\and collaboration approach \\for multi-cell FL.\end{tabular}                             & \begin{tabular}[c]{@{}c@{}}The convergence speed \\has not been \\further discussed.\textcolor[rgb]{0.204,0.208,0.255}{}\end{tabular}                           \\ 
\hline
\multirow{3}{*}{\begin{tabular}[c]{@{}c@{}}Joint \\Optimization\end{tabular}} & \multirow{3}{*}{\begin{tabular}[c]{@{}c@{}}Minimize \\optimality gap\end{tabular}}                      & \begin{tabular}[c]{@{}c@{}}Discrete enumeration \\algorithm~\cite{fan2021joint}\end{tabular}                                                    & \begin{tabular}[c]{@{}c@{}}The algorithm is capable of\\~adapting to~multiple\\~local update approaches.\end{tabular}                                   & \begin{tabular}[c]{@{}c@{}}The tightening of \\the problem solution space \\leads to suboptimality.\textcolor[rgb]{0.204,0.208,0.255}{}\end{tabular}            \\ 
\cline{3-5}
                                                                              &                                                                                                         & \begin{tabular}[c]{@{}c@{}}SDR technique after \\problem transformation~\cite{guo2022joint}\end{tabular}                                        & \begin{tabular}[c]{@{}c@{}}The algorithm simultaneously \\considers the impact of\\~uplink and downlink noise.\end{tabular}                             & \begin{tabular}[c]{@{}c@{}}Suboptimality is introduced~when \\converting the discrete problem \\into a continuous form.\end{tabular}                            \\ 
\cline{3-5}
                                                                              &                                                                                                         & \begin{tabular}[c]{@{}c@{}}Sorting algorithm after \\problem transformation \cite{10001136}\end{tabular}                                        & \begin{tabular}[c]{@{}c@{}}The computational complexity\\~ is reduced through \\problem transformation.\end{tabular}                                    & \begin{tabular}[c]{@{}c@{}}The problem-solving process\\~is heuristic in nature.\end{tabular}                                                                   \\
\hline
\end{tabular}
    }
\end{table*}

\section{Multi-Antenna OTA-FL}

Recently, there has been a growing interest in exploring more complex scenarios involving multiple antennas at the transmitters and/or receivers in OTA-FL. In such cases, joint optimization of transmitter and receiver beamforming, along with user selection and power control, becomes crucial to achieve enhanced performance. Compared to the SISO OTA-FL systems, the design of OTA-FL systems with multiple antennas is still at an early stage, offering significant opportunities for further development and exploration. 

\begin{figure}[!h]
	\centering
	\includegraphics[width=1\linewidth]{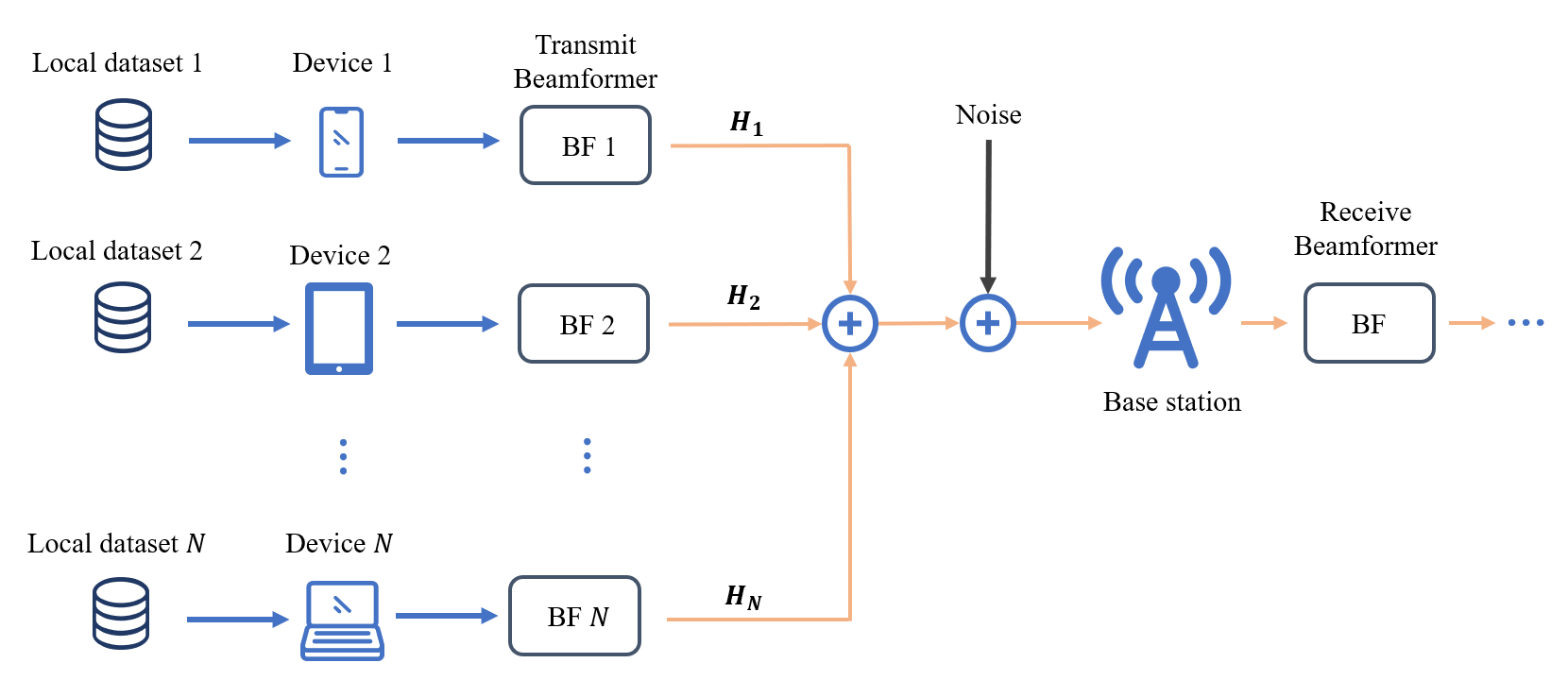}
	\caption{An illustration of multi-antenna OTA-FL system. To fully utilize the benefits of multiple antennas, a joint design approach is commonly adopted to optimize user selection, power control and beamforming vectors.}
	\label{fig:System Model of MIMO Aircomp}
\end{figure}

Given the strong interdependence among variables in multi-antenna OTA-FL systems, existing studies predominantly adopt a joint design approach, aiming to optimize variables such as user selection, power control, and beamforming vectors, which are highly coupled, as illustrated in Fig. \ref{fig:System Model of MIMO Aircomp}.
These joint design approaches are motivated by the need to exploit the benefits of multiple antennas in OTA-FL systems, such as improved spectral efficiency and enhanced communication reliability. By jointly optimizing user selection, power control, and beamforming, the performance of multi-antenna OTA-FL systems can be significantly improved, thereby unlocking the potential of utilizing multiple antennas effectively for distributed learning tasks. In Table \ref{mimo table}, we summarize  the efforts made by existing studies in exploring MIMO OTA-FL systems and summarize their respective optimization objectives and methods, as well as their strengths and limitations.

In multi-antenna OTA-FL systems, the prevailing approach for optimization is centered around minimizing MSE. For instance, in \cite{9321510}, the authors tackle the problem of optimizing users' transmit beamforming and the BS's receive beamforming to minimize the MSE of the received signals. To address the challenges posed by highly coupled variables, they propose an alternating minimization algorithm that jointly optimizes the transmit and receive beamforming vectors. However, recent studies suggest that exploring alternative communication-learning representation objectives may offer a more promising approach than solely focusing on MSE. While MSE minimization remains the mainstream practice, researchers are starting to recognize the potential benefits of adopting different optimization objectives.

In the context of MIMO OTA-FL systems, \cite{8708985} considers the optimization of wireless power transfer (WPT). They reformulate the non-convex MSE minimization problem into two nested sub-problems, making it tractable and enabling efficient solutions. 
Another approach is presented in \cite{8468002}, where equalization and channel feedback techniques are designed to improve OTA computation capabilities. The objective is to minimize computation MSE, and the authors propose an equalization strategy based on differential geometry and a channel feedback mechanism to facilitate OTA computation functions. 
Furthermore, there are studies that explore the integration of OTA with other systems. In \cite{10014666}, OTA is combined with an integrated sensing and communication (ISAC) system to leverage the advantages of both sensing and OTA computation. The optimization objective in this case is to minimize radar sensing MSE while maintaining sensing accuracy.

In the context of IoT networks, the authors of \cite{8807380} propose a MIMO OTA scheme for networks with clustered multi-antenna sensors and a receive array at the base station. They introduce an optimal receive beamformer, known as the decomposed aggregation beamformer (DAB), which utilizes a decomposed architecture to reduce the channel dimension and perform joint equalization. Additionally, they propose a low-latency channel feedback framework that leverages OTA to enable simultaneous channel feedback from the sensors.

To address signaling overhead and power consumption concerns, the authors of \cite{9107137} present a joint optimization scheme that minimizes MSE through the optimization of transmit beamformers and hybrid combiners. This scheme achieves local stable convergence by dividing time slots and optimizing two subproblems: short-term and long-term optimizations. 
In \cite{yang2020federated}, a joint approach incorporating user selection and receiving beamforming design is utilized to achieve fast model aggregation. This approach employs a difference-of-convex-functions (DC) representation to improve sparsity and accurately detect the fixed-rank constraint.
Similarly, an efficient user selection scheme is achieved in \cite{10146443} by maximizing the number of selected devices and minimizing the aggregation error. A greedy method based on matching pursuit is designed to reduce computational complexity and preserve training performance of the DC method achieved in \cite{yang2020federated}.

While the majority of existing studies in multi-antenna OTA-FL focus on MSE minimization, there are only a few studies that explore minimizing the optimality gap. For example, the authors of \cite{zhong2021over} propose a strategy to jointly optimize transceiver beamforming and user selection by minimizing the upper bound. They develop an algorithm based on an alternating optimization framework with low complexity, which helps address the optimization problem and alleviate the straggler problem by aligning the upload gradients at the BS.

\begin{table*}[hbt] 
    \centering
\renewcommand\arraystretch{1.2}  
\caption{Summary of the existing studies on Multi-Antenna OTA-FL systems}
\label{mimo table}
\resizebox{\textwidth}{!}{
\begin{tabular}{|c|c|c|c|} \hline
\textbf{Objective function}                                                                 & \textbf{Method}                                                                                                        & \textbf{Pros}                                                                                                                                 & \textbf{Cons}                                                                                                                                                                   \\ \hline
\multirow{6}{*}{\begin{tabular}[c]{@{}c@{}}Minimize the MSE \\ of received signal\end{tabular}} & SCA-based optimization~algorithm \cite{9321510}                                                                        & \begin{tabular}[c]{@{}c@{}}It can achieve performance \\close to that of \\the fully-digital beamforming.\end{tabular}                        & \begin{tabular}[c]{@{}c@{}}SCA approach is employed, \\which leads to \\ suboptimal solutions.\end{tabular}                                                            \\ \cline{2-4}
                                                                                            & \begin{tabular}[c]{@{}c@{}}Equivalence between original problem \\ and nested sub-problems \cite{8708985}\end{tabular} & \begin{tabular}[c]{@{}c@{}}The algorithm can lead to \\a global optimal solution.\end{tabular}                                                & \begin{tabular}[c]{@{}c@{}}The computational complexity \\increases sharply \\with the number of antennas.\end{tabular}                                                         \\ \cline{2-4}
                                                                                            & Grassmann manifold approximation \cite{8468002}                                                                        & \begin{tabular}[c]{@{}c@{}}The algorithm effectively \\reduces computational workload \\compared to SDR approach.\end{tabular}                & \begin{tabular}[c]{@{}c@{}}Due to the limitations on \\nomographic functions, \\OTA channel feedback~ \\is not applicable to traditional \\multi-user MIMO cases.\end{tabular}  \\ \cline{2-4}
                                                                                            & SDR based algorithm \cite{10014666}                                                                                     & \begin{tabular}[c]{@{}c@{}}The system is capable of \\simultaneously fulfilling \\communication and sensing.\end{tabular}                     & \begin{tabular}[c]{@{}c@{}}The embedding of sensing~\\results in a partial degradation of \\communication performance.\end{tabular}                                             \\ \cline{2-4}
                                                                                            & One-dimensional search algorithm \cite{8807380}                                                                        & \begin{tabular}[c]{@{}c@{}}It achieves \\low-latency channel feedback \\based on OTA characteristics.\end{tabular}                            & \begin{tabular}[c]{@{}c@{}}The original optimization problem \\needs several approximation, \\which causes~suboptimality.\end{tabular}                                          \\ \cline{2-4}
                                                                                            & \begin{tabular}[c]{@{}c@{}}Approximate equivalence \\ of solving subproblems \cite{9107137}\end{tabular}               & \begin{tabular}[c]{@{}c@{}}The scheme is designed to \\operate without \\explicit S-CSI requirements.\end{tabular}                            & \begin{tabular}[c]{@{}c@{}}The heuristic algorithm \\can only obtain~\\a local optimal solution.\end{tabular}                                                                   \\ \hline
\multirow{2}{*}{\begin{tabular}[c]{@{}c@{}}Minimize MSE and\\ maximize quantity of devices\end{tabular}} & \begin{tabular}[c]{@{}c@{}}DC programming\\ after problem transformation \cite{yang2020federated}\end{tabular} & \begin{tabular}[c]{@{}c@{}}Effective user selection \\ is achieved by obtaining \\ the low-rank sparse structure \\ of the problem.\end{tabular} & \begin{tabular}[c]{@{}c@{}}The SDR-based solution \\ results in \\ high computation complexity.\end{tabular} \\ \cline{2-4} 
                                                                                                         & \begin{tabular}[c]{@{}c@{}}Greedy algorithm \\ based on matching pursuit \cite{10146443}\end{tabular}          & \begin{tabular}[c]{@{}c@{}}The computational complexity\\ is effectively reduced.\end{tabular}                               & \begin{tabular}[c]{@{}c@{}}The loss function \\ needs to be strongly convex\\ which leads to certain limitations.\end{tabular}                 \\ \hline
\begin{tabular}[c]{@{}c@{}}Minimize \\ optimality gap\end{tabular}                          & \begin{tabular}[c]{@{}c@{}}Alternating optimization \\ and fractional programming \cite{zhong2021over}\end{tabular}    & \begin{tabular}[c]{@{}c@{}}The scheme enables to \\alleviate system straggler issues.\end{tabular}                                            & \begin{tabular}[c]{@{}c@{}}The strong convexity assumption \\for FL loss functions \\has certain limitations.\end{tabular}                                                      \\ \hline
\end{tabular}
}
\end{table*}

\section{RIS-assisted OTA-FL}

\begin{figure}[!h]
	\centering
	\includegraphics[width=1\linewidth]{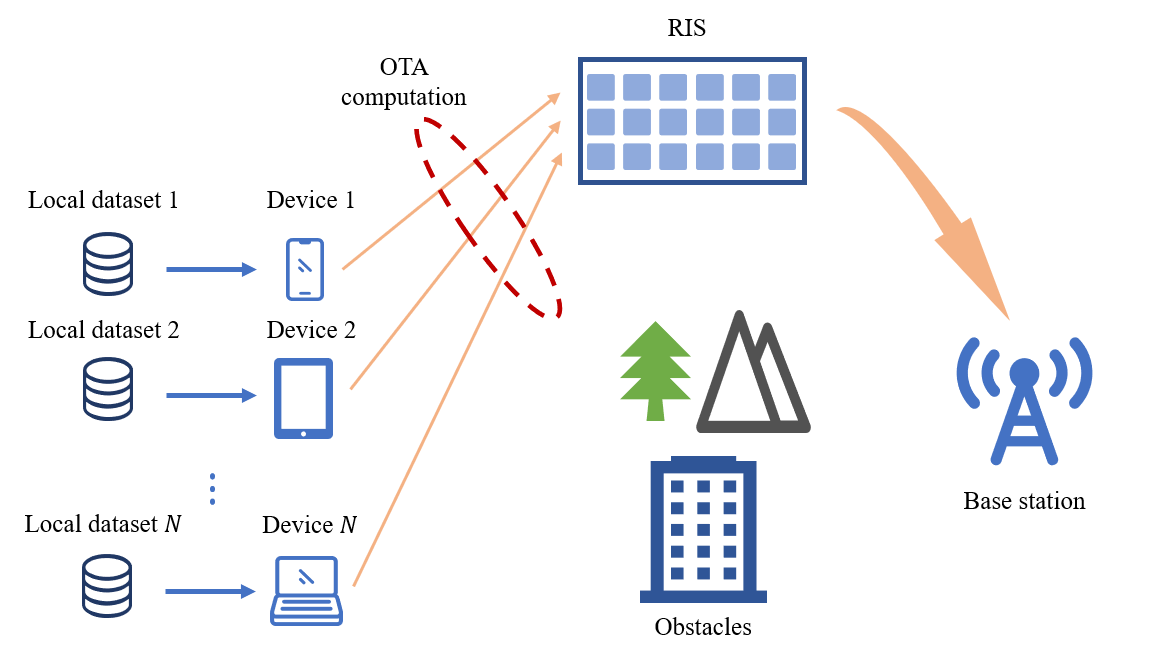}
	\caption{An illustration of RIS-assisted OTA-FL system, where the RIS alters the propagation direction of reflected signals by manipulating the phase of incident signals, thereby enhancing the received signal strength at the BS.}
	\label{fig:Sketch of RIS OTA-FL}
\end{figure}

In the context of the envisioned development of 6G, the integration of RIS technology is gaining significant attention and is considered a promising solution for OTA-FL systems. RIS, as a cost-effective passive device, consists of multiple reflective elements capable of adjusting the phase shift of incoming signals, consequently altering the propagation direction of the reflected signals~\cite{10146001,9999559,9217160,9580328}, as illustrated in Fig.~\ref{fig:Sketch of RIS OTA-FL}. This unique capability of RIS enables the superimposition of the direct link signal, effectively mitigating signal energy attenuation resulting from obstacles such as buildings~\cite{10021680}.

RIS holds great potential for improving the performance of OTA-FL systems by overcoming signal propagation challenges. By strategically deploying RIS in the wireless environment, it becomes possible to enhance signal strength and quality, leading to improved communication efficiency and higher reliability~\cite{9806300,9373634,9707727}. The use of RIS technology can effectively address issues related to signal attenuation and multipath fading, which are common obstacles in wireless communications. Additionally, RIS can offer flexibility in optimizing signal propagation paths, thus enabling better coverage and reduced interference in OTA-FL systems.

Several studies have focused on incorporating RIS into OTA-FL frameworks, as delineated in the following. We provide a detailed summary of OTA-FL systems assisted by RISs in Table \ref{ris table}, where the optimization objectives, variables, and methods of the systems, as well as their strengths and limitations, are presented accordingly.

\subsection{RIS-Assisted OTA-FL in Single-Antenna Systems}

In OTA-FL systems incorporating RISs, comprehensive optimization of multiple variables is necessary, including RIS configurations, user selection, and power control. However, these optimization problems can be challenging due to their non-convex nature and the close coupling of variables.

In \cite{9502547}, the authors propose the use of RISs to mitigate the impact of fading channels and enable reliable model aggregation. It is important to note that increasing the number of reflecting elements in the RIS can aggregate more devices for training and improve performance. However, this also leads to increased channel estimation overhead. Thus, a co-design of power control and resource distribution is required to strike a balance. Similarly, in \cite{9451567}, the trade-off is investigated between communication and learning, highlighting the essentiality of RIS deployment. The authors propose an algorithm that jointly optimizes user scheduling, receiver beamforming vectors, and RIS phase shifts to achieve the desired performance.

In order to address performance bottlenecks under poor propagation channel conditions, OTA-FL systems with RISs are designed in \cite{9546764} and \cite{9797713}. These works focus on the co-design of transceivers and RIS phase shifts to minimize aggregation errors. In \cite{9525086}, a mixed-timescale penalty-dual-decomposition (MTPDD) algorithm is proposed to reduce signaling overhead caused by a large number of RIS elements. The algorithm jointly minimizes the MSE of computation over time while mitigating signaling overhead.
The potential of OTA-FL in multi-cell networks is explored in \cite{9928626} through the utilization of RISs. The authors propose an alternating minimization scheme to optimize receiving beamforming vectors and RIS phase shifts jointly. 
Furthermore, in~\cite{9829187}, the concept of a double-RIS (DRIS) OTA-FL system is introduced, where a pair of RISs are placed at the users' and BS ends, respectively. This configuration addresses the problem of unavailable direct links due to obstacles. It is demonstrated that the MSE performance of the DRIS system surpasses that of single RIS systems when a substantial number of RIS elements and receiving antennas are deployed.

While it may be challenging to obtain the global optimum in the works mentioned above, significant performance gains have been achieved through the rational use of RISs and the design of optimization algorithms. These studies contribute to the exploration and advancement of OTA-FL systems, leveraging the benefits of RIS technology.

\subsection{RIS-Assisted OTA-FL in Multi-Antenna Systems}

Introducing multiple antennas in RIS-assisted OTA-FL systems provides increased degrees of freedom but also leads to a larger number of variables that need to be optimized. 
In \cite{9414785}, the authors present a RIS-assisted OTA-FL framework with multiple antennas at the BS. They formulate an energy minimization problem that aims to jointly optimize user selection, phase shifts, decoding vectors, and power control. To handle the complexity of the problem, it is decomposed into several subproblems and solved iteratively. 
In \cite{9815198}, user terminals are equipped with multiple antennas to facilitate concurrent model transmission over a millimeter wave (mmWave) network. The objective is to minimize transmission distortion. Thus, the joint optimization of RIS phase shifts and beamforming vectors is considered to achieve this goal. 
Similarly, in~\cite{9682077}, a DRIS-assisted OTA-FL system with multiple antennas is proposed to enhance the channel quality. The objective is to minimize MSE by jointly optimizing receiving beamforming vectors, denoising factor, power control, and passive beamforming design. By considering all these variables together, the system performance can be improved.

In these works, the authors recognize the importance of optimizing multiple antennas in RIS-assisted OTA-FL systems. By jointly considering variables, such as user selection, phase shifts, beamforming vectors, power control, and passive beamforming design, the performance of the systems can be enhanced and the objectives can be achieved more effectively.

A critical challenge in RIS-assisted OTA-FL systems is the assumption of perfect instantaneous CSI in existing studies. However, obtaining accurate CSI can be computationally expensive or even infeasible, particularly in scenarios where the system is mobile and the channel exhibits high dynamics. This limitation poses a significant hurdle in optimizing the performance of such systems.

The integration of RIS technology in OTA-FL systems represents a promising avenue for research and development. It has the potential to revolutionize wireless communications by mitigating signal attenuation and improving overall system performance. As the exploration of RIS in OTA-FL continues to expand, further advancements and optimizations are expected to unlock its full potential in realizing efficient and reliable next-generation wireless networks.

\begin{table*}[hbt]
    \centering
    \renewcommand\arraystretch{1.2}  
    \caption{Comparative summary of the existing studies on RIS-assisted OTA-FL systems}
    \label{ris table}
    \resizebox{\textwidth}{!}{
        \begin{tabular}{|c|c|c|c|c|c|} 
\hline
\textbf{Antenna Setting}        & \textbf{Joint Design}                                                                                                          & \textbf{Objective function}                                                                 & \textbf{Method}                                                                                & \textbf{Pros}                                                                                                                                                               & \textbf{Cons}                                                                                                                                    \\ 
\hline
\multirow{7}{*}{Single-antenna} & \multirow{2}{*}{\begin{tabular}[c]{@{}c@{}}User selection, \\RIS phase shifts \\ and beamformer at BS\end{tabular}}            & \begin{tabular}[c]{@{}c@{}}Maximize the\\number of \\ scheduled users\end{tabular}          & \begin{tabular}[c]{@{}c@{}}Alternating \\minimization \\ algorithm \cite{9502547}\end{tabular} & \begin{tabular}[c]{@{}c@{}}The device selection \\process achieves \\performance close to\\~that of an exhaustive \\approach.\end{tabular}                                  & \begin{tabular}[c]{@{}c@{}}The algorithm \\achieves~better \\performance but\\~at the cost of \\higher computational \\complexity.\end{tabular}  \\ 
\cline{3-6}
                                &                                                                                                                                & \begin{tabular}[c]{@{}c@{}}Minimize \\ optimality gap\end{tabular}                          & \begin{tabular}[c]{@{}c@{}}SCA-based \\optimization \\ algorithm \cite{9451567}\end{tabular}   & \begin{tabular}[c]{@{}c@{}}The proposed \\algorithm has \\low computational \\complexity.\end{tabular}                                                                      & \begin{tabular}[c]{@{}c@{}}The assumption \\is made that \\the loss function \\is strongly convex.\end{tabular}                                  \\ 
\cline{2-6}
                                & \begin{tabular}[c]{@{}c@{}}Tranmit scalar, \\RIS phase shifts, \\ beamforming vectors\\and denoising factor at BS\end{tabular} & \multirow{5}{*}{\begin{tabular}[c]{@{}c@{}}Minimize the MSE \\ of received signal\end{tabular}} & \begin{tabular}[c]{@{}c@{}}Alternating \\minimization \\ algorithm~\cite{9546764}\end{tabular} & \begin{tabular}[c]{@{}c@{}}The algorithm \\is able to \\reduce computational \\time while achieving \\similar performance.\textcolor[rgb]{0.216,0.255,0.318}{}\end{tabular} & \begin{tabular}[c]{@{}c@{}}The approximation of \\the problem leads to \\suboptimal solutions.\end{tabular}                                      \\ 
\cline{2-2}\cline{4-6}
                                & \begin{tabular}[c]{@{}c@{}}Transceiver design \\and RIS phase shifts\end{tabular}                                              &                                                                                             & \begin{tabular}[c]{@{}c@{}}Alternating \\minimization \\ algorithm \cite{9797713}\end{tabular} & \begin{tabular}[c]{@{}c@{}}The subproblems \\have closed-form \\solutions after \\problem decomposition.\end{tabular}                                                       & \begin{tabular}[c]{@{}c@{}}Specific trade-offs~\\regarding the \\number of RISs have \\not been provided.\end{tabular}                           \\ 
\cline{2-2}\cline{4-6}
                                & \begin{tabular}[c]{@{}c@{}}Transmit power, \\RIS passive beamforming\\ and beamformer at BS\end{tabular}                       &                                                                                             & \begin{tabular}[c]{@{}c@{}}Penalty-dual-decomposition \\ algorithm~\cite{9525086}\end{tabular} & \begin{tabular}[c]{@{}c@{}}The system performance\\~is able to match the \\theoretical upper bound.\end{tabular}                                                            & \begin{tabular}[c]{@{}c@{}}The case of large-scale \\fading has not \\been mentioned.\end{tabular}                                               \\ 
\cline{2-2}\cline{4-6}
                                & \begin{tabular}[c]{@{}c@{}}RIS phase shifts \\and beamformer at BS\end{tabular}                                                &                                                                                             & \begin{tabular}[c]{@{}c@{}}Alternating \\minimization \\ algorithm~\cite{9928626}\end{tabular} & \begin{tabular}[c]{@{}c@{}}The~ scheme achieves \\more efficient multi-cell \\interference management.\end{tabular}                                                         & \begin{tabular}[c]{@{}c@{}}The computational \\complexity of the \\algorithm has not \\been mentioned.\end{tabular}                              \\ 
\cline{2-2}\cline{4-6}
                                & \begin{tabular}[c]{@{}c@{}}Transmit scalar, \\RIS passive beamforming\\ and beamformer at BS\end{tabular}                      &                                                                                             & \begin{tabular}[c]{@{}c@{}}Alternating \\minimization \\ algorithm~\cite{9829187}\end{tabular} & \begin{tabular}[c]{@{}c@{}}Theoretical proofs \\demonstrate the \\performance advantages \\of DRIS over a single RIS.\end{tabular}                                          & \begin{tabular}[c]{@{}c@{}}The proposed algorithm \\exhibits weaker \\performance compared to \\SDR-based approaches.\end{tabular}               \\ 
\hline
\multirow{3}{*}{Multi-antenna}  & \begin{tabular}[c]{@{}c@{}}User selection, \\RIS phase shifts, \\ decoding vector, \\and power control\end{tabular}            & \begin{tabular}[c]{@{}c@{}}Minimize \\ power-delay \\product\end{tabular}                   & \begin{tabular}[c]{@{}c@{}}Problem \\decomposition~\cite{9414785}\end{tabular}                 & \begin{tabular}[c]{@{}c@{}}The algorithm is able to \\utilize RISs to reduce \\energy consumption.\end{tabular}                                                             & \begin{tabular}[c]{@{}c@{}}The algorithm cannot \\guarantee optimality\\~through iterative solving\\~of individual subproblems.\end{tabular}     \\ 
\cline{2-6}
                                & \begin{tabular}[c]{@{}c@{}}Transmission distortion, \\ beamforming vectors, \\and RIS phase shifts\end{tabular}                & \multirow{2}{*}{\begin{tabular}[c]{@{}c@{}}Minimize the MSE \\ of received signal\end{tabular}} & \begin{tabular}[c]{@{}c@{}}Alternating \\minimization \\ algorithm~\cite{9815198}\end{tabular} & \begin{tabular}[c]{@{}c@{}}The algorithm does \\not require the \\assumption of \\strongly convex \\loss functions.\end{tabular}                                            & \begin{tabular}[c]{@{}c@{}}The computational\\complexity~of the \\algorithm has not~\\been mentioned.\end{tabular}                               \\ 
\cline{2-2}\cline{4-6}
                                & \begin{tabular}[c]{@{}c@{}}Beamforming vectors, \\denoising factor, \\ power control \\and passive beamforming\end{tabular}    &                                                                                             & \begin{tabular}[c]{@{}c@{}}Alternating \\minimization \\ algorithm \cite{9682077}\end{tabular} & \begin{tabular}[c]{@{}c@{}}Closed-form \\solution structures\\~can be obtained for \\part of the problems.\end{tabular}                                                     & \begin{tabular}[c]{@{}c@{}}No theoretical \\justification is \\provided for the \\superiority of DRIS \\over a single RIS.\end{tabular}          \\
\hline
\end{tabular}
    }
\end{table*}

\section{Trust, Security and Privacy of  OTA-FL}

FL has made significant strides in improving privacy compared to centralized learning, as it enables local processing of raw data. However, the baseline OTA-FL model still lacks a formal guarantee of trust, security, and privacy \cite{kairouz2021advances}. While these terms are often used interchangeably in existing literature, it is important to highlight their distinct meanings.

Trust ensures that processes operate in expected ways, providing assurance to stakeholders. Security, on the other hand, focuses on protecting data from unauthorized access or alteration, including guarding against Byzantine attacks. Privacy is concerned with preventing the disclosure of private information during interactions with other entities \cite{safeguarding}.

To address the challenges and ensure trust, security, and privacy in OTA-FL, various mechanisms have been proposed and investigated. Researchers have explored different approaches to develop OTA-FL systems that preserve these crucial aspects. These mechanisms aim to establish a robust framework that instills confidence in the system's operations, safeguards against unauthorized access and malicious attacks, and protects the privacy of sensitive information.
By emphasizing the distinctions between trust, security, and privacy, researchers can better understand the multifaceted nature of OTA-FL systems and design comprehensive solutions. By integrating these mechanisms into the OTA-FL model, it becomes possible to establish a trustworthy, secure, and privacy-preserving framework that meets the requirements of various stakeholders involved in the collaborative learning process. Table \ref{fig:DP-mechanism} presents a summary of the existing studies on the trust, security, and privacy of OTA-FL, where the design objectives and security or privacy guarantees are systematically reviewed.

\subsection{Trust and Integrity}

Trust is a concept that reflects the level of control and confidence that an entity has in another entity. It can also be seen as an outcome resulting from advancements in achieving security and privacy objectives \cite{tr933ec,10109152}. In the context of FL, trust plays a significant role. For example, enterprises must trust network operators when granting consent for the collection of their data in FL. However, the abundance of valuable private data introduces concerns regarding algorithms that aim to predict specific business states.

The lack of trust among participants in FL remains an ongoing challenge. In an effort to address this issue, FLChain is proposed in \cite{8905038}. FLChain is a decentralized, publicly auditable, and healthy FL ecosystem that incorporates trust and incentives. It aims to create an environment where participants can trust the system and each other. However, to date, no research specifically focuses on establishing trust in the field of trustworthy OTA-FL.

In order to foster trust in OTA-FL systems, it is crucial to explore mechanisms that enhance transparency, accountability, and verifiability. These mechanisms can include decentralized architectures, public auditing, and incentives for participants. By incorporating these elements into the design and implementation of OTA-FL frameworks, researchers can work towards establishing trust among the involved entities and ensuring the reliability and security of the collaborative learning process.

\begin{figure}[hbt]
	\centering
	\includegraphics[width=1\linewidth]{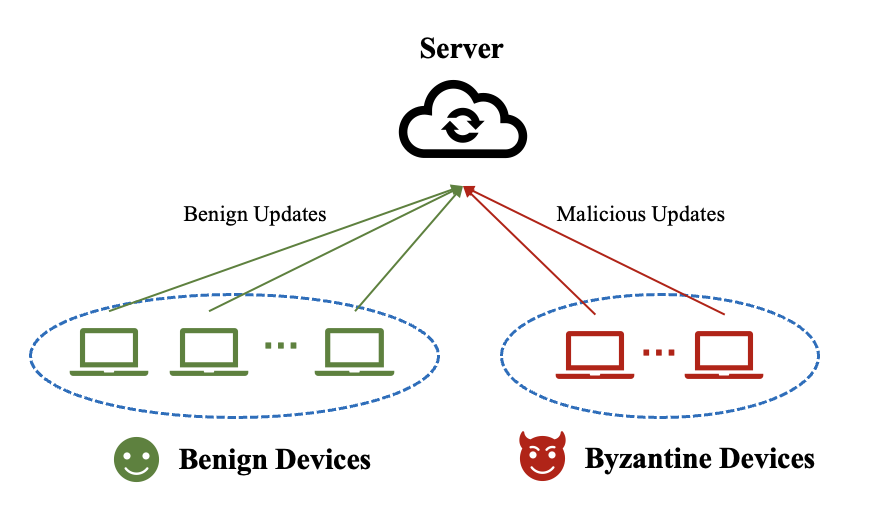}
	\caption{An illustration of the OTA-FL framework with a server, benign devices, and Byzantine devices, where the Byzantine devices launch attacks to contaminate the benign devices through the global model aggregation at the server~\cite{huang2021byzantine}.}
	\label{fig:Byzantine-mechanism}
\end{figure}

\subsection{Security}

The security concerns associated with OTA-FL primarily revolve around the potential for poisoning attacks, where malicious participants aim to compromise the federated learning process. These attacks can be classified into two main categories:
\begin{itemize}
\item \textit{Data Poisoning}: One example of a data poisoning attack is the label-flipping attack, as described in \cite{label-flipping}. In this type of attack, the adversary maintains the same features in their training sample but flips the corresponding label to a different class. The adversary aims to corrupt the training process and influence the learned model by introducing such maliciously labeled data~\cite{10129254}.
\item 
\textit{Model Poisoning}: Adversaries in OTA-FL have the ability to manipulate local model updates before transmitting them to the server. This manipulation can lead to various outcomes, such as causing misclassifications or implanting hidden backdoors~\cite{9870690} within the global model. It has been observed that in FL scenarios, model poisoning attacks tend to be more effective compared to data poisoning attacks \cite{bhagoji2019analyzing}.
\end{itemize}
These poisoning attacks pose significant security risks to OTA-FL, as they can undermine the integrity and reliability of the federated learning process. Mitigating these threats requires robust security measures and techniques to detect and prevent malicious behavior from compromising the OTA-FL system.

The simplicity of OTA-FL procedures can render the learning process vulnerable to intentional poisoning attacks by adversaries. In recent years, there has been growing interest in Byzantine attacks, where malicious devices aim to disrupt FL convergence or steer it towards a poisoned model, as illustrated in Fig. \ref{fig:Byzantine-mechanism}. Therefore, it is crucial to design secure OTA-FL systems that can effectively counteract these attacks. Currently, only a few works specifically address Byzantine attacks in the context of OTA-FL, which are described below.

For instance, in \cite{sifaou2022over}, the authors propose ROTAF, a novel transmission and aggregation approach that enhances the robustness of OTA-FL against Byzantine attacks in the case of I.I.D. data distribution. In ROTAF, participating devices are divided into different groups during each global training round, and each group is assigned a separate transmit time slot. The local updates from different groups are aggregated using geometric median aggregation. When dealing with non-I.I.D. data, a resampling step is performed before applying geometric median aggregation. The authors provide theoretical convergence analysis of ROTAF under both I.I.D. and non-I.I.D. data assumptions, demonstrating its ability to converge to within a range of the optimum at a linear rate when the number of groups exceeds twice the number of attacks. Numerical results show that ROTAF exhibits robustness against various forms of Byzantine attacks compared to the basic averaging OTA-FL approach.

In \cite{huang2021byzantine}, the authors focus on reducing the computation complexity of Byzantine-resilient OTA-FL. They address the challenge posed by the complexity of solving a convex problem in the commonly-used geometric median aggregation approach when the model parameter dimension is large. To overcome this, they adopt the improved Weiszfeld algorithm to calculate the smoothed geometric median. By leveraging the additive structure of the Weiszfeld algorithm, which can be combined with OTA, they propose a secure aggregation approach that is jointly designed for computation and communication in OTA-FL. 
In \cite{fan2022bev}, the authors highlight the limitations of the widely-used channel inversion method in OTA-FL for defending against Byzantine attacks. They introduce a novel policy called best effort voting (BEV), which integrates local stochastic gradient descent (SGD) to ensure safe aggregation. BEV allows users to transmit their local gradients with maximum power, maximizing the deterrent effect on OTA-FL convergence. The authors analyze the convergence of BEV and demonstrate its superiority over popular channel inversion methods, particularly under strong adversarial environments.

These works represent important contributions to addressing Byzantine attacks in OTA-FL, providing techniques and mechanisms to enhance the security and resilience of the learning process against malicious behavior.

\begin{figure}
	\centering
	\includegraphics[width=1\linewidth]{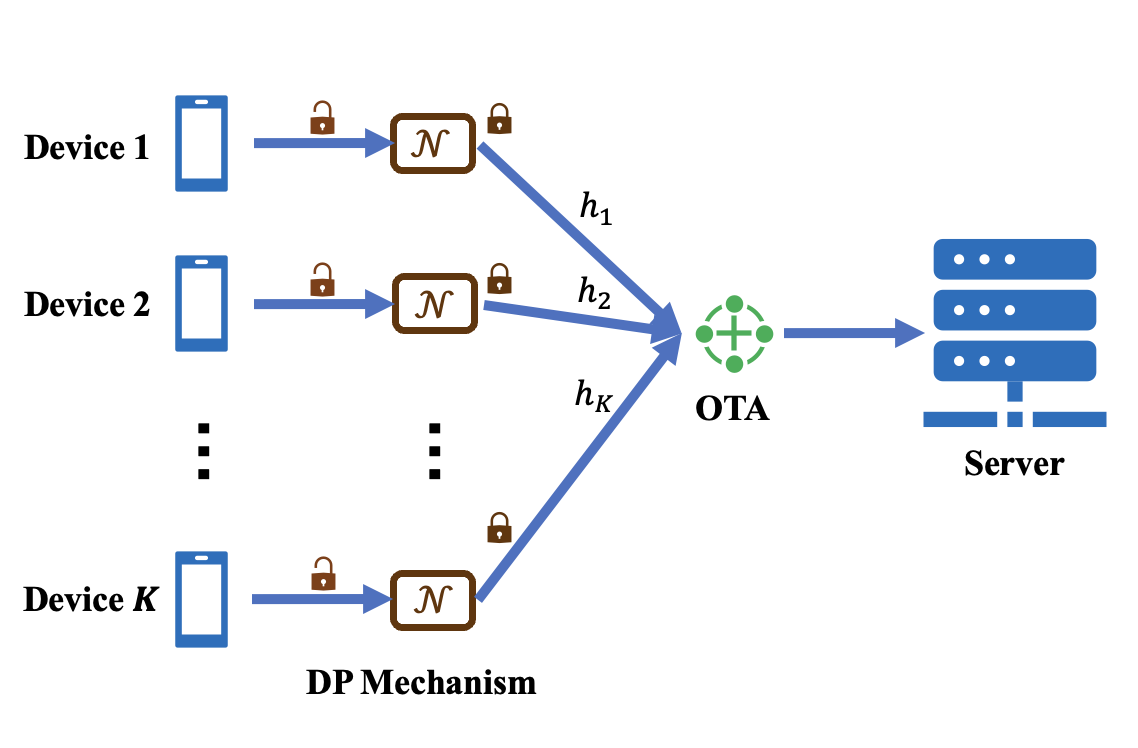}
	\caption{An illustration of the OTA-FL framework with DP, where the communication between devices and the server must adhere to DP requirements for each user and be obfuscated with artificial Laplacian or Gaussian noises~\cite{9174426}.}
	\label{fig:DP-mechanism}
\end{figure}

\begin{table*}[hbt]
\centering
\renewcommand\arraystretch{1.2}  
\caption{Summary of the existing consideration of Trust, Secure and Privacy in OTA-FL systems}
\resizebox{\textwidth}{!}{
\begin{tabular}{|c|c|c|c|c|} 
\hline
\textbf{Category}                  & \textbf{Counteracting}                                                                                                     & \textbf{Method}                               & \textbf{Guarantee}                                                                                 & \textbf{Description}                                                                                                                      \\ 
\hline
\textbf{Trust}                     & \multicolumn{4}{c|}{This area is important but current research is limited.}                                                                                                                                                                                                                                                                                                                                                \\ 
\hline
\multirow{3}{*}{\textbf{Security}} & \multirow{3}{*}{\begin{tabular}[c]{@{}c@{}}Data poisoning \\and model poisoning \\(e.g., backdoor attacks)\end{tabular}}   & \multirow{2}{*}{Geometric median aggregation} & \multirow{3}{*}{Byzantine-resilience}                                                              & \begin{tabular}[c]{@{}c@{}}Introduce Weiszfeld algorithm \\to lower computation complexity. \cite{huang2021byzantine}\end{tabular}       \\ 
\cline{5-5}
                                   &                                                                                                                            &                                               &                                                                                                    & \begin{tabular}[c]{@{}c@{}}Group devices and assign \\transmit slots before aggregation.~\cite{sifaou2022over}\end{tabular}              \\ 
\cline{3-3}\cline{5-5}
                                   &                                                                                                                            & Best effort voting                            &                                                                                                    & \begin{tabular}[c]{@{}c@{}}Allow devices to transmit with maximum power \\to defend Byzantine attacks.~\cite{fan2022bev}\end{tabular}    \\ 
\hline
\multirow{10}{*}{\textbf{Privacy}} & \multirow{10}{*}{\begin{tabular}[c]{@{}c@{}}Membership inference, \\property inference\\~and model inversion\end{tabular}} & \multirow{2}{*}{Using inherent channel noise} & \multirow{8}{*}{Differential privacy}                                                              & \begin{tabular}[c]{@{}c@{}}Minimize the optimality gap with privacy\\~obtained ``for free" from channel.~\cite{9252950}\end{tabular}   \\ 
\cline{5-5}
                                   &                                                                                                                            &                                               &                                                                                                    & \begin{tabular}[c]{@{}c@{}}Develop a power control strategy \\to preserve DP from channel noise.~\cite{9322199}\end{tabular}          \\ 
\cline{3-3}\cline{5-5}
                                   &                                                                                                                            & \multirow{6}{*}{Adding artificial noise}      &                                                                                                    & \begin{tabular}[c]{@{}c@{}}Combine the local gradients linearly \\with artificial Gaussian noise.~\cite{9174426}\end{tabular}            \\ 
\cline{5-5}
                                   &                                                                                                                            &                                               &                                                                                                    & \begin{tabular}[c]{@{}c@{}}Propose misaligned power allocation \\to enhance the system SNR.~\cite{9810958}\end{tabular}               \\ 
\cline{5-5}
                                   &                                                                                                                            &                                               &                                                                                                    & \begin{tabular}[c]{@{}c@{}}Add spatially correlated perturbation noise \\to the local updates.~\cite{9929413}\end{tabular}               \\ 
\cline{5-5}
                                   &                                                                                                                            &                                               &                                                                                                    & \begin{tabular}[c]{@{}c@{}}Distribute the noise generating process \\to defend pilot attacks.~\cite{9413624}\end{tabular}                \\ 
\cline{5-5}
                                   &                                                                                                                            &                                               &                                                                                                    & \begin{tabular}[c]{@{}c@{}}Propose a privacy-preserving variant \\of the second-order method ADMM.~\cite{9427268}\end{tabular}           \\ 
\cline{5-5}
                                   &                                                                                                                            &                                               &                                                                                                    & \begin{tabular}[c]{@{}c@{}}Reduce the update dimension \\and enhance the communication efficiency.~\cite{9685320}\end{tabular}           \\ 
\cline{3-5}
                                   &                                                                                                                            & \multirow{2}{*}{Adding artificial noise}      & \multirow{2}{*}{\begin{tabular}[c]{@{}c@{}}Differential privacy \\secure aggregation\end{tabular}} & \begin{tabular}[c]{@{}c@{}}Add pairwise cancellable noise \\to thwart external eavesdroppers.~\cite{xue2022over}\end{tabular}            \\ 
\cline{5-5}
                                   &                                                                                                                            &                                               &                                                                                                    & \begin{tabular}[c]{@{}c@{}}Assign part of devices to send noise \\to degrade the SNR of eavesdroppers.~\cite{yan2022toward}\end{tabular}  \\
\hline
\end{tabular}
}
\end{table*}

\subsection{Privacy}
While FL provides a decentralized approach to model training without the need to share local data with the BS or other participants, it is important to note that FL is not immune to privacy concerns~\cite{yu2023obfuscating}. Recent research has demonstrated that FL is susceptible to inference attacks~\cite{10073536}, which can potentially lead to the recovery of local training data. This vulnerability arises because the model or gradient updates obtained from local data can unintentionally reveal additional information about the underlying features in the data that were not intended to be disclosed. In the context of OTA-FL, attacks that exploit this privacy vulnerability can be categorized as follows: 
\begin{itemize}
    \item \textit{Membership Inference}\cite{membership-inference}: Adversary could test whether a specific data partition of a device has been utilized for training a model.
    On the FourSquare location dataset, the authors show that a malicious server can $99\%$ convincingly tell if particular location metadata is used to train a classifier.
    \item \textit{Property Inference}\cite{property-inference}: Adversary could test if a specific partition of data with certain properties is contained in the data of a device. It is significant to emphasize that this property may not be directly associated with the primary objective.
    \item \textit{Model Inversion}\cite{model-inversion}: Adversary could reconstruct an input sample of the training data of a device based on the local updates. One sample of the training dataset is generated, which should be private with a generative adversarial network (GAN).
\end{itemize}

It is crucial to design private-by-design OTA-FL systems that are inherently against inference attacks. To this end, differential privacy (DP), one of the perturbation-based methods, has been adopted as a standard solution for preserving privacy in OTA-FL, as illustrated in Fig. \ref{fig:DP-mechanism}.
DP serves as a robust standard for ensuring privacy in distributed systems \cite{abadi2016deep}. 
A randomized mechanism $\mathcal{N}:\mathcal{D}\rightarrow\mathcal{K}$, where $\mathcal{D}$ denotes domain and $\mathcal{K}$ denotes range, is of ($\epsilon$,$\delta$)-DP, on condition that for arbitrary two adjacent inputs $i,i^{'} \in\mathcal{D}$ and for an arbitrary subset of outputs $O\subseteq\mathcal{K}$ it holds~\cite{dwork2014algorithmic}:
\begin{equation}\label{DP-definition}
    \text{Pr}[\mathcal{N}(i)\in O]\leq e^\epsilon \text{Pr}[\mathcal{N}(i^{'})\in O]+\delta
\end{equation}
For suitably small constants $\epsilon$ and $\delta$, it is statistically impossible for an adversary to violate privacy because of the indistinguishability of neighboring datasets $l$ and $l^{'}$. 

The preservation of privacy in OTA-FL can be achieved through the introduction of artificial noise to the local updates~\cite{9069945}. For instance, in \cite{9174426}, the authors consider OTA-FL with local DP over flat-fading Gaussian channels, where privacy-constrained artificial Gaussian noise is linearly combined with the local gradients during transmission. Analytical results show a tradeoff between privacy and convergence for certain loss functions, indicating that the training error decreases as the device count $K$ increases. It is also demonstrated that the privacy level per user decreases at a rate of $\mathcal{O}(1/\sqrt{K})$ compared to orthogonal transmission, where privacy leakage remains constant. However, a limitation of this scheme is that the device constrains the system signal-to-noise ratio (SNR) with the weakest channel, which can degrade learning accuracy. To address this issue, the MPA-DPFL scheme with misaligned power allocation is proposed in \cite{9810958}. The scheme suggests allocating power misaligned when a device's channel gain falls below a certain threshold, as opposed to the aligned manner used in \cite{9174426}.

Several works have considered more comprehensive approaches to introducing additional noise to local updates. In~\cite{9929413}, the authors propose a novel DP-based method that adds spatially correlated perturbations to the local updates at each device. Compared to traditional DP-based methods that employ uncorrelated noise, this approach achieves higher learning accuracy while preserving defense ability. Additionally, in~\cite{9413624}, it is emphasized that CSI at the devices is crucial for transmission and ensuring a DP-based privacy guarantee. The proposed distributed noise generation process is resilient against pilot attacks manipulated by a malicious server and the failure of transmitting nodes. 

Another technique, presented in \cite{9427268}, is a so-called A-FADMM algorithm based on the alternating direction method of multipliers (ADMM). By adding a random variable to the local update and multiplying it with a random fading gain, this method preserves the privacy of both the local model trajectory and gradient trajectory. To enhance communication efficiency, the authors in \cite{9685320} propose the differentially private random projection FedSGD scheme, which reduces the dimension of local updates while preserving privacy.

An alternative to artificial noises is that the inherent channel noise can also be leveraged to preserve privacy. In \cite{9252950}, the authors demonstrate that privacy can be obtained ``for free" from the channel noise when the privacy constraint level is below a certain threshold that decreases with SNR. It is also highlighted that actively assigning additional power to perturb local gradients is generally suboptimal in OTA-FL scenarios. To address this, a dynamic power control strategy is optimized under power and privacy constraints to minimize the optimality gap. Similarly, in \cite{9322199}, the inherent receiver noise is harnessed to preserve DP against inference attacks, and a novel power control strategy is introduced. The analysis shows that the received SNR is primarily influenced by the number of devices when aiming for a higher level of privacy.

Recent research has been focused on the joint design of secure and private OTA-FL systems. In the work by Xue et al. \cite{xue2022over}, they consider a scenario where an external attacker equipped with a directional antenna eavesdrops on RF signals from the devices, which poses a more potent attack model that existing approaches struggle to handle. To address this challenge, the authors propose the introduction of pairwise suppressible random, artificial noises. These noises are used to obfuscate private local model parameters and thwart external eavesdroppers. This design can be seen as an integration of DP and secure aggregation at the physical layer for OTA-FL. 
In addition, Yan et al.~\cite{yan2022toward} propose a secure and private OTA-FL framework, which utilizes noise to preserve privacy and security guarantees. The framework employs DP and MSE-security as the metrics. Specifically, a subset of devices is designated to send Gaussian artificial noise with the aim of degrading the SNR of potential eavesdroppers. To mitigate the impact of noise on learning accuracy, a channel-weighted post-processing mechanism is introduced. Moreover, the authors propose a scheduling algorithm based on the branch-and-bound concept with low complexity. This algorithm ensures the security of the system and the privacy of user data stored on the server.

\section{Lessons Learned, Open Challenges, and Future Directions}
As OTA-FL in wireless environments continues to gain attention, researchers are actively working to tackle the associated challenges and improve system performance. However, there are still several open questions and directions for further research in this area. Some of these challenging questions are discussed in the following. 

\subsection{Aggregation Distortion}
OTA-FL systems face the challenge of distortion introduced by channel fading, noise, and transceiver filtering, which can degrade the quality of the received summation signal \cite{9666041}, \cite{9895450}. Minimizing this distortion has been a persistent challenge in OTA-FL. Coded OTA can help reduce distortion but adds complexity to the system with coding and decoding processes. On the other hand, uncoded OTA requires an advanced transceiver design to achieve optimal amplitude alignment and combat interference. Both approaches necessitate improved design techniques to enhance the training performance of OTA-FL systems.

\subsection{Stringent Synchronization Requirement}
The assumption of perfect signal synchronization at the receiving end has been commonly made in most existing OTA-FL studies \cite{Our-Paper}. However, this assumption becomes increasingly challenging to achieve in scenarios with large system sizes and high heterogeneity. While some efforts have been made to address this challenge through robust design techniques, such as those proposed in \cite{6573232}, effectively implementing synchronization in complex network environments remains an important and unresolved area that requires further investigation. Overcoming the synchronization challenge is crucial to ensure the reliable and efficient operation of OTA-FL in real-world wireless systems.

\subsection{Data Heterogeneity }
In different OTA-FL scenarios, user data often have different distributions. It is necessary to consider different, non-I.I.D. settings when testing the performance of different algorithms to ensure a robust design. For example, the authors of \cite{9835537} define several non-I.I.D. distribution policies to serve as benchmarks. Meanwhile, the severely unbalanced distribution of data often leads to the gradient importance of different users, which makes some users' updates submerged in receiver noises. To this end, an adequate design of aggregation weights under non-I.I.D. distributions is a vital direction in the future. 

As a matter of fact, OTA-FL is particularly susceptible to unbalanced volumes of training data among different users. This is because local models trained based on significantly larger amounts of local data are typically weighted higher. In the context of OTA-FL, this means the local models would be delivered with much higher received powers at the server (or BS). A near-far effect could occur, leading to the loss of local modes trained based on smaller amounts of data and delivered with lower transmit powers. 

\subsection{Secure and Trustworthy OTA-FL}
While efforts have been made to protect user data by submerging it within the overlay signal, and mitigate the risk of information theft, there remains an inherent risk of leakage~\cite{9446488}. To ensure reliable aggregation computation and minimize aggregation errors caused by active eavesdroppers in different environments, it is crucial to design robust transceiver policies. These policies should effectively counteract the presence of eavesdroppers and maintain the integrity and privacy of the aggregated data.

In addition to addressing security concerns, building users' confidence in local data aggregation is paramount. To achieve this, it is essential to develop reliable algorithms and establish appropriate metrics in trustworthy FL. Researchers should focus their attention on creating robust and verifiable algorithms that instill trust in the aggregation process and provide transparent metrics to assess the reliability and accuracy of the aggregated results. By emphasizing the importance of trustworthy FL and dedicating efforts to its development, researchers can contribute to enhancing users' confidence in the integrity and privacy of their data.

\subsection{Other Challenges}
In addition, it is crucial to consider the impact of inherent channel fading and additive noise within OTA-FL systems on the convergence upper bound. These factors play a significant role in system performance and should be appropriately captured by a communication-learning metric. Identifying and defining a metric that effectively captures the influence of channel fading and noise is an essential direction for further research in OTA-FL.

Moreover, the implementation of the right to data erasure within the OTA-FL framework presents challenges. The right to data erasure allows users to request the removal of their data from a dataset or model under specific circumstances to protect their privacy \cite{qu2023learn}. However, handling this ``unlearning" situation in OTA-FL is not straightforward. Permanently removing user data from the system can lead to significant performance losses, particularly in non-I.I.D. data distributions. Therefore, developing techniques that address data erasure while minimizing the impact on system performance is a challenging aspect that requires further exploration in OTA-FL research.

\section{Conclusion}
This paper has provided a comprehensive overview of the latest studies on the emerging OTA-FL technique. We first categorized OTA-FL systems under different system settings, including single-antenna and multiple-antenna OTA-FL systems, as well as the consideration of RISs. The design objectives and optimization tools were analyzed. Next, we delineated the trust, security, and privacy aspects of OTA-FL systems, provided corresponding performance evaluation metrics, and unveiled critical concerns needed to promote better system design. Additionally, we highlighted the challenges faced by OTA-FL and suggested future research directions. Challenges to be holistically addressed include  model distortion under channel fading, the ineffective OTA aggregation of local models trained on substantially unbalanced data, and the limited accessibility and verifiability of individual local models. 

% \section*{Acknowledgments}

\bibliographystyle{ieeetr}
\bibliography{arxiv_references}

\begin{thebibliography}{10}

\bibitem{10002946}
K.~Li, Y.~Cui, W.~Li, T.~Lv, X.~Yuan, S.~Li, W.~Ni, M.~Simsek, and F.~Dressler,
  ``When {Internet of Things} meets {M}etaverse: Convergence of physical and
  cyber worlds,'' {\em IEEE Internet Things J.}, vol.~10, no.~5,
  pp.~4148--4173, 2023.

\bibitem{peltonen20206g}
E.~Peltonen, M.~Bennis, M.~Capobianco, M.~Debbah, A.~Ding,
  F.~Gil-Casti{\~n}eira, M.~Jurmu, T.~Karvonen, M.~Kelanti, A.~Kliks, {\em
  et~al.}, ``6{G} white paper on edge intelligence,'' {\em \rm arXiv preprint
  arXiv:2004.14850}, 2020.

\bibitem{eldar2022machine}
Y.~C. Eldar, A.~Goldsmith, D.~G{\"u}nd{\"u}z, and H.~V. Poor, {\em Machine
  {L}earning and {W}ireless {C}ommunications}.
\newblock Cambridge University Press, 2022.

\bibitem{10118940}
J.~Zheng, K.~Li, N.~Mhaisen, W.~Ni, E.~Tovar, and M.~Guizani, ``Federated
  learning for online resource allocation in mobile edge computing: A deep
  reinforcement learning approach,'' in {\em Proc. IEEE Wireless Commun.
  Networking Conf. (WCNC)}, pp.~1--6, 2023.

\bibitem{10000870}
L.~Pu, Q.~Cui, X.~L. I, B.~Zhao, W.~Ni, M.~Ai, and X.~Tao, ``Federated
  learning-based heterogeneous load prediction and slicing for {5G} systems and
  beyond,'' in {\em Proc. IEEE Global Commun. Conf. (GLOBECOM)}, pp.~166--172,
  2022.

\bibitem{9920736}
K.~Li, W.~Ni, Y.~Emami, and F.~Dressler, ``Data-driven flight control of
  {Internet-of-Drones} for sensor data aggregation using multi-agent deep
  reinforcement learning,'' {\em IEEE Wireless Commun.}, vol.~29, no.~4,
  pp.~18--23, 2022.

\bibitem{9779339}
J.~Zheng, K.~Li, N.~Mhaisen, W.~Ni, E.~Tovar, and M.~Guizani, ``Exploring
  deep-reinforcement-learning-assisted federated learning for online resource
  allocation in privacy-preserving {EdgeIoT},'' {\em IEEE Internet Things J.},
  vol.~9, no.~21, pp.~21099--21110, 2022.

\bibitem{mcmahan2017communication}
B.~McMahan, E.~Moore, D.~Ramage, S.~Hampson, and B.~A.~y. Arcas,
  ``{Communication-Efficient Learning of Deep Networks from Decentralized
  Data},'' in {\em Proc. Int. Conf. Artif. Intell. Stat. (AISTATS)},
  pp.~1273--1282, 2017.

\bibitem{10123399}
W.~Li, T.~Lv, Y.~Cao, W.~Ni, and M.~Peng, ``Multi-carrier {NOMA}-empowered
  wireless federated learning with optimal power and bandwidth allocation,''
  {\em IEEE Trans. Wireless Commun.}, 2023, early access.

\bibitem{9880724}
X.~Chen, Z.~Li, W.~Ni, X.~Wang, S.~Zhang, S.~Xu, and Q.~Pei, ``Two-phase deep
  reinforcement learning of dynamic resource allocation and client selection
  for hierarchical federated learning,'' in {\em Proc. IEEE Int. Conf. Commun.
  China (ICCC)}, pp.~518--523, 2022.

\bibitem{4305404}
B.~Nazer and M.~Gastpar, ``Computation over multiple-access channels,'' {\em
  IEEE Trans. Inf. Theory}, vol.~53, no.~10, pp.~3498--3516, 2007.

\bibitem{9415623}
D.~C. Nguyen, M.~Ding, P.~N. Pathirana, A.~Seneviratne, J.~Li, and H.~V.~Poor,
  ``Federated learning for {I}nternet of {T}hings: A comprehensive survey,''
  {\em IEEE Commun. Surv. Tutorials}, vol.~23, no.~3, pp.~1622--1658, 2021.

\bibitem{6557530}
M.~Goldenbaum, H.~Boche, and S.~Stańczak, ``Harnessing interference for analog
  function computation in wireless sensor networks,'' {\em IEEE Trans. Signal
  Process.}, vol.~61, no.~20, pp.~4893--4906, 2013.

\bibitem{9796935}
B.~Luo, W.~Xiao, S.~Wang, J.~Huang, and L.~Tassiulas, ``Tackling system and
  statistical heterogeneity for federated learning with adaptive client
  sampling,'' in {\em Proc. IEEE Int. Conf. Comput. Commun. (INFOCOM)},
  pp.~1739--1748, 2022.

\bibitem{8870236}
G.~Zhu, Y.~Wang, and K.~Huang, ``Broadband analog aggregation for low-latency
  federated edge learning,'' {\em IEEE Trans. Wireless Commun.}, vol.~19,
  no.~1, pp.~491--506, 2020.

\bibitem{9605599}
Y.~Sun, S.~Zhou, Z.~Niu, and D.~Gündüz, ``Dynamic scheduling for over-the-air
  federated edge learning with energy constraints,'' {\em IEEE J. Sel. Areas
  Commun.}, vol.~40, no.~1, pp.~227--242, 2022.

\bibitem{9844173}
G.~Shi, S.~Guo, J.~Ye, N.~Saeed, and S.~Dang, ``Multiple parallel federated
  learning via over-the-air computation,'' {\em IEEE Open J. Commun. Soc.},
  vol.~3, pp.~1252--1264, 2022.

\bibitem{10038617}
J.~Du, B.~Jiang, C.~Jiang, Y.~Shi, and Z.~Han, ``Gradient and channel aware
  dynamic scheduling for over-the-air computation in federated edge learning
  systems,'' {\em IEEE J. Sel. Areas Commun.}, vol.~41, no.~4, pp.~1035--1050,
  2023.

\bibitem{cao2022transmission}
X.~Cao, G.~Zhu, J.~Xu, and S.~Cui, ``Transmission power control for
  over-the-air federated averaging at network edge,'' {\em IEEE J. Sel. Areas
  Commun.}, vol.~40, no.~5, pp.~1571--1586, 2022.

\bibitem{9606731}
X.~Cao, G.~Zhu, J.~Xu, Z.~Wang, and S.~Cui, ``Optimized power control design
  for over-the-air federated edge learning,'' {\em IEEE J. Sel. Areas Commun.},
  vol.~40, no.~1, pp.~342--358, 2022.

\bibitem{9843892}
Y.~Zou, Z.~Wang, X.~Chen, H.~Zhou, and Y.~Zhou, ``Knowledge-guided learning for
  transceiver design in over-the-air federated learning,'' {\em IEEE Trans.
  Wireless Commun.}, vol.~22, no.~1, pp.~270--285, 2023.

\bibitem{9791337}
Z.~Wang, Y.~Zhou, Y.~Shi, and W.~Zhuang, ``Interference management for
  over-the-air federated learning in multi-cell wireless networks,'' {\em IEEE
  J. Sel. Areas Commun.}, vol.~40, no.~8, pp.~2361--2377, 2022.

\bibitem{9780892}
S.~Jing and C.~Xiao, ``Federated learning via over-the-air computation with
  statistical channel state information,'' {\em IEEE Trans. Wireless Commun.},
  vol.~21, no.~11, pp.~9351--9365, 2022.

\bibitem{10039388}
X.~Yu, B.~Xiao, W.~Ni, and X.~Wang, ``Optimal power control for over-the-air
  federated edge learning using statistical channel knowledge,'' in {\em Proc.
  IEEE Int. Conf. Wireless Commun. Signal Process. (WCSP)}, pp.~232--237, 2022.

\bibitem{9076343}
T.~Sery and K.~Cohen, ``On analog gradient descent learning over multiple
  access fading channels,'' {\em IEEE Trans. Signal Process.}, vol.~68,
  pp.~2897--2911, 2020.

\bibitem{Our-Paper}
X.~Yu, B.~Xiao, W.~Ni, and X.~Wang, ``Optimal adaptive power control for
  over-the-air federated edge learning under fading channels,'' {\em IEEE
  Trans. Commun.}, pp.~1--1, 2023.

\bibitem{fan2021joint}
X.~Fan, Y.~Wang, Y.~Huo, and Z.~Tian, ``Joint optimization of communications
  and federated learning over the air,'' {\em IEEE Trans. Wireless Commun.},
  vol.~21, no.~6, pp.~4434--4449, 2022.

\bibitem{guo2022joint}
W.~Guo, R.~Li, C.~Huang, X.~Qin, K.~Shen, and W.~Zhang, ``Joint device
  selection and power control for wireless federated learning,'' {\em IEEE J.
  Sel. Areas Commun.}, vol.~40, no.~8, pp.~2395--2410, 2022.

\bibitem{10001136}
C.~Chen, Y.-H. Chiang, H.~Lin, J.~C. Lui, and Y.~Ji, ``Energy harvesting aware
  client selection for over-the-air federated learning,'' in {\em Proc. IEEE
  Global Commun. Conf. (GLOBECOM)}, pp.~5069--5074, 2022.

\bibitem{9321510}
X.~Zhai, X.~Chen, J.~Xu, and D.~W. Kwan~Ng, ``Hybrid beamforming for massive
  {MIMO} over-the-air computation,'' {\em IEEE Trans. Commun.}, vol.~69, no.~4,
  pp.~2737--2751, 2021.

\bibitem{8708985}
X.~Li, G.~Zhu, Y.~Gong, and K.~Huang, ``Wirelessly powered data aggregation for
  {IoT} via over-the-air function computation: Beamforming and power control,''
  {\em IEEE Trans. Wireless Commun.}, vol.~18, no.~7, pp.~3437--3452, 2019.

\bibitem{8468002}
G.~Zhu and K.~Huang, ``{MIMO} over-the-air computation for high-mobility
  multimodal sensing,'' {\em IEEE Internet Things J.}, vol.~6, no.~4,
  pp.~6089--6103, 2019.

\bibitem{10014666}
X.~Li, F.~Liu, Z.~Zhou, G.~Zhu, S.~Wang, K.~Huang, and Y.~Gong, ``Integrated
  sensing, communication, and computation over-the-air: {MIMO} beamforming
  design,'' {\em IEEE Trans. Wireless Commun.}, pp.~1--1, 2023.

\bibitem{8807380}
D.~Wen, G.~Zhu, and K.~Huang, ``Reduced-dimension design of {MIMO} over-the-air
  computing for data aggregation in clustered {IoT} networks,'' {\em IEEE
  Trans. Wireless Commun.}, vol.~18, no.~11, pp.~5255--5268, 2019.

\bibitem{9107137}
X.~Chen, A.~Liu, and M.-J. Zhao, ``High-mobility multi-modal sensing for {IoT}
  network via {MIMO} aircomp: A mixed-timescale optimization approach,'' {\em
  IEEE Commun. Lett.}, vol.~24, no.~10, pp.~2295--2299, 2020.

\bibitem{yang2020federated}
K.~Yang, T.~Jiang, Y.~Shi, and Z.~Ding, ``Federated learning via over-the-air
  computation,'' {\em IEEE Trans. Wireless Commun.}, vol.~19, no.~3,
  pp.~2022--2035, 2020.

\bibitem{10146443}
A.~Bereyhi, A.~Vagollari, S.~Asaad, R.~R. Müller, W.~Gerstacker, and H.~V.
  Poor, ``Device scheduling in over-the-air federated learning via matching
  pursuit,'' {\em IEEE Trans. Signal Process.}, vol.~71, pp.~2188--2203, 2023.

\bibitem{zhong2021over}
C.~Zhong, H.~Yang, and X.~Yuan, ``Over-the-air multi-task federated learning
  over {MIMO} interference channel,'' {\em \rm arXiv preprint
  arXiv:2112.13603}, 2021.

\bibitem{10146001}
S.~Hu, X.~Yuan, W.~Ni, X.~Wang, and A.~Jamalipour, ``{RIS}-assisted jamming
  rejection and path planning for {UAV}-borne {IoT} platform: A new deep
  reinforcement learning framework,'' {\em IEEE Internet Things J.}, 2023,
  early access.

\bibitem{9999559}
R.~Saleem, W.~Ni, M.~Ikram, and A.~Jamalipour,
  ``Deep-reinforcement-learning-driven secrecy design for
  intelligent-reflecting-surface-based {6G-IoT} networks,'' {\em IEEE Internet
  Things J.}, vol.~10, no.~10, pp.~8812--8824, 2023.

\bibitem{9217160}
Y.~Cao, T.~Lv, and W.~Ni, ``Intelligent reflecting surface aided multi-user
  {mmWave} communications for coverage enhancement,'' in {\em Proc. IEEE Int.
  Symp. Pers. Indoor Mob. Radio Commun. (PIMRC)}, pp.~1--6, 2020.

\bibitem{9580328}
C.~Sun, W.~Ni, Z.~Bu, and X.~Wang, ``Energy minimization for intelligent
  reflecting surface-assisted mobile edge computing,'' in {\em Proc. IEEE Int.
  Conf. Commun. China (ICCC)}, pp.~254--259, 2021.

\bibitem{10021680}
X.~Yuan, S.~Hu, W.~Ni, R.~P. Liu, and X.~Wang, ``Joint user, channel,
  modulation-coding selection, and {RIS} configuration for jamming resistance
  in multiuser {OFDMA} systems,'' {\em IEEE Trans. Commun.}, vol.~71, no.~3,
  pp.~1631--1645, 2023.

\bibitem{9806300}
Y.~Cao, T.~Lv, and W.~Ni, ``Two-timescale optimization for intelligent
  reflecting surface-assisted {MIMO} transmission in fast-changing channels,''
  {\em IEEE Trans. Wireless Commun.}, vol.~21, no.~12, pp.~10424--10437, 2022.

\bibitem{9373634}
Y.~Cao, T.~Lv, Z.~Lin, and W.~Ni, ``Delay-constrained joint power control, user
  detection and passive beamforming in intelligent reflecting surface-assisted
  uplink {mmWave} system,'' {\em IEEE Trans. Cognit. Commun. Networking},
  vol.~7, no.~2, pp.~482--495, 2021.

\bibitem{9707727}
C.~Sun, W.~Ni, Z.~Bu, and X.~Wang, ``Energy minimization for intelligent
  reflecting surface-assisted mobile edge computing,'' {\em IEEE Trans.
  Wireless Commun.}, vol.~21, no.~8, pp.~6329--6344, 2022.

\bibitem{9502547}
Z.~Wang, J.~Qiu, Y.~Zhou, Y.~Shi, L.~Fu, W.~Chen, and K.~B. Letaief,
  ``Federated learning via intelligent reflecting surface,'' {\em IEEE Trans.
  Wireless Commun.}, vol.~21, no.~2, pp.~808--822, 2022.

\bibitem{9451567}
H.~Liu, X.~Yuan, and Y.-J.~A. Zhang, ``Reconfigurable intelligent surface
  enabled federated learning: A unified communication-learning design
  approach,'' {\em IEEE Trans. Wireless Commun.}, vol.~20, no.~11,
  pp.~7595--7609, 2021.

\bibitem{9546764}
W.~Fang, Y.~Jiang, Y.~Shi, Y.~Zhou, W.~Chen, and K.~B. Letaief, ``Over-the-air
  computation via reconfigurable intelligent surface,'' {\em IEEE Trans.
  Commun.}, vol.~69, no.~12, pp.~8612--8626, 2021.

\bibitem{9797713}
W.~Zhang, J.~Xu, W.~Xu, X.~You, and W.~Fu, ``Worst-case design for {RIS}-aided
  over-the-air computation with imperfect {CSI},'' {\em IEEE Commun. Lett.},
  vol.~26, no.~9, pp.~2136--2140, 2022.

\bibitem{9525086}
X.~Zhai, G.~Han, Y.~Cai, and L.~Hanzo, ``Beamforming design based on two-stage
  stochastic optimization for {RIS}-assisted over-the-air computation
  systems,'' {\em IEEE Internet Things J.}, vol.~9, no.~7, pp.~5474--5488,
  2022.

\bibitem{9928626}
J.~Li, Y.~Wu, and Y.~Shi, ``Reconfigurable intelligent surface assisted
  over-the-air computation in multi-cell networks,'' in {\em Proc. IEEE Int.
  Mediterr. Conf. Commun. Networking (MeditCom)}, pp.~43--48, 2022.

\bibitem{9829187}
X.~Zhai, G.~Han, Y.~Cai, and L.~Hanzo, ``Joint beamforming aided over-the-air
  computation systems relying on both {BS}-side and user-side reconfigurable
  intelligent surfaces,'' {\em IEEE Trans. Wireless Commun.}, vol.~21, no.~12,
  pp.~10766--10779, 2022.

\bibitem{9414785}
Y.~Hu, M.~Chen, M.~Chen, Z.~Yang, M.~Shikh-Bahaei, H.~V. Poor, and S.~Cui,
  ``Energy minimization for federated learning with {IRS}-assisted over-the-air
  computation,'' in {\em Proc. IEEE Int. Conf. Acoust. Speech Signal Process.
  (ICASSP)}, pp.~3105--3109, 2021.

\bibitem{9815198}
L.~Hu, Z.~Wang, H.~Zhu, and Y.~Zhou, ``{RIS}-assisted over-the-air federated
  learning in millimeter wave {MIMO} networks,'' {\em J. Commun. Inf.
  Networks}, vol.~7, no.~2, pp.~145--156, 2022.

\bibitem{9682077}
J.~Li, M.~Fu, Y.~Zhou, and Y.~Shi, ``Double-{RIS} assisted over-the-air
  computation,'' in {\em Proc. IEEE GC Wkshps}, pp.~1--6, 2021.

\bibitem{kairouz2021advances}
P.~Kairouz, H.~B. McMahan, B.~Avent, A.~Bellet, M.~Bennis, A.~N. Bhagoji,
  K.~Bonawitz, Z.~Charles, G.~Cormode, R.~Cummings, {\em et~al.}, ``Advances
  and open problems in federated learning,'' {\em Found. Trends Mach. Learn.},
  vol.~14, no.~1--2, pp.~1--210, 2021.

\bibitem{safeguarding}
C.~Ma, J.~Li, M.~Ding, H.~H. Yang, F.~Shu, T.~Q.~S. Quek, and H.~V. Poor, ``On
  safeguarding privacy and security in the framework of federated learning,''
  {\em IEEE Network}, vol.~34, no.~4, pp.~242--248, 2020.

\bibitem{tr933ec}
N.~Robinson, L.~Valeri, J.~Cave, T.~Starkey, H.~Graux, S.~Creese, and
  P.~Hopkins, {\em Understanding the {I}mplications for {S}ecurity, {P}rivacy
  and {T}rust}, pp.~23--27.
\newblock RAND Corporation, 2011.

\bibitem{10109152}
S.~A. Siddiqui, A.~Mahmood, Q.~Z. Sheng, H.~Suzuki, and W.~Ni, ``Trust in
  vehicles: Toward context-aware trust and attack resistance for the {I}nternet
  of {V}ehicles,'' {\em IEEE Trans. Intell. Transp. Syst.}, pp.~1--15, 2023,
  early access.

\bibitem{8905038}
X.~Bao, C.~Su, Y.~Xiong, W.~Huang, and Y.~Hu, ``F{L}{C}hain: A blockchain for
  auditable federated learning with trust and incentive,'' in {\em Proc. Int.
  Conf. Big Data Comput. Commun. (BIGCOM)}, pp.~151--159, 2019.

\bibitem{huang2021byzantine}
S.~Huang, Y.~Zhou, T.~Wang, and Y.~Shi, ``Byzantine-resilient federated machine
  learning via over-the-air computation,'' in {\em Proc. IEEE ICC Wkshps},
  pp.~1--6, IEEE, 2021.

\bibitem{label-flipping}
C.~Fung, C.~J. Yoon, and I.~Beschastnikh, ``Mitigating sybils in federated
  learning poisoning,'' {\em \rm arXiv preprint arXiv:1808.04866}, 2018.

\bibitem{10129254}
Y.~Wang, T.~Li, S.~Li, X.~Yuan, and W.~Ni, ``New adversarial image detection
  based on sentiment analysis,'' {\em IEEE Trans. Neural Networks Learn.
  Syst.}, pp.~1--15, 2023, early access.

\bibitem{9870690}
Y.~Wang, M.~Zhao, S.~Li, X.~Yuan, and W.~Ni, ``Dispersed pixel
  perturbation-based imperceptible backdoor trigger for image classifier
  models,'' {\em IEEE Trans. Inf. Forensics Secur.}, vol.~17, pp.~3091--3106,
  2022.

\bibitem{bhagoji2019analyzing}
A.~N. Bhagoji, S.~Chakraborty, P.~Mittal, and S.~Calo, ``Analyzing federated
  learning through an adversarial lens,'' in {\em Int. Conf. Mach. Learn.},
  pp.~634--643, PMLR, 2019.

\bibitem{sifaou2022over}
H.~Sifaou and G.~Y. Li, ``Over-the-air federated learning under {B}yzantine
  attacks,'' {\em \rm arXiv preprint arXiv:2205.02949}, 2022.

\bibitem{fan2022bev}
X.~Fan, Y.~Wang, Y.~Huo, and Z.~Tian, ``{BEV-SGD}: Best effort voting {SGD}
  against byzantine attacks for analog-aggregation-based federated learning
  over the air,'' {\em IEEE Internet Things J.}, vol.~9, no.~19,
  pp.~18946--18959, 2022.

\bibitem{9174426}
M.~Seif, R.~Tandon, and M.~Li, ``Wireless federated learning with local
  differential privacy,'' in {\em Proc. IEEE Int. Symp. Inf. Theory (ISIT)},
  pp.~2604--2609, 2020.

\bibitem{9252950}
D.~Liu and O.~Simeone, ``Privacy for free: Wireless federated learning via
  uncoded transmission with adaptive power control,'' {\em IEEE J. Sel. Areas
  Commun.}, vol.~39, no.~1, pp.~170--185, 2021.

\bibitem{9322199}
Y.~Koda, K.~Yamamoto, T.~Nishio, and M.~Morikura, ``Differentially private
  aircomp federated learning with power adaptation harnessing receiver noise,''
  in {\em Proc. IEEE Global Commun. Conf. (GLOBECOM)}, pp.~1--6, 2020.

\bibitem{9810958}
N.~Yan, K.~Wang, C.~Pan, and K.~K. Chai, ``Private federated learning with
  misaligned power allocation via over-the-air computation,'' {\em IEEE Commun.
  Lett.}, vol.~26, no.~9, pp.~1994--1998, 2022.

\bibitem{9929413}
J.~Liao, Z.~Chen, and E.~G. Larsson, ``Over-the-air federated learning with
  privacy protection via correlated additive perturbations,'' in {\em Proc.
  Annu. Allerton Conf. Commun. Control Comput. (Allerton)}, pp.~1--8, 2022.

\bibitem{9413624}
B.~Hasırcıoğlu and D.~Gündüz, ``Private wireless federated learning with
  anonymous over-the-air computation,'' in {\em Proc. IEEE Int. Conf. Acoust.
  Speech Signal Process. (ICASSP)}, pp.~5195--5199, 2021.

\bibitem{9427268}
A.~Elgabli, J.~Park, C.~B. Issaid, and M.~Bennis, ``Harnessing wireless
  channels for scalable and privacy-preserving federated learning,'' {\em IEEE
  Trans. Commun.}, vol.~69, no.~8, pp.~5194--5208, 2021.

\bibitem{9685320}
A.~Sonee, S.~Rini, and Y.~Huang, ``Wireless federated learning with limited
  communication and differential privacy,'' in {\em Proc. IEEE Global Commun.
  Conf. (GLOBECOM)}, pp.~01--06, 2021.

\bibitem{xue2022over}
X.~Xue, M.~K. Hasan, S.~Yu, L.~N. Kandel, and M.~Song, ``Over-the-air federated
  learning with enhanced privacy,'' {\em \rm arXiv preprint arXiv:2212.11486},
  2022.

\bibitem{yan2022toward}
N.~Yan, K.~Wang, K.~Zhi, C.~Pan, K.~K. Chai, and H.~V. Poor, ``Toward secure
  and private over-the-air federated learning,'' {\em \rm arXiv preprint
  arXiv:2210.07669}, 2022.

\bibitem{yu2023obfuscating}
G.~Yu, X.~Wang, C.~Sun, P.~Yu, W.~Ni, and R.~P. Liu, ``Obfuscating the dataset:
  Impacts and applications,'' {\em ACM Trans. Intell. Syst. Technol.}, 2023.

\bibitem{10073536}
X.~Yuan, W.~Ni, M.~Ding, K.~Wei, J.~Li, and H.~V. Poor, ``Amplitude-varying
  perturbation for balancing privacy and utility in federated learning,'' {\em
  IEEE Trans. Inf. Forensics Secur.}, vol.~18, pp.~1884--1897, 2023.

\bibitem{membership-inference}
R.~Shokri, M.~Stronati, C.~Song, and V.~Shmatikov, ``Membership inference
  attacks against machine learning models,'' in {\em Proc. IEEE Symp. Secur.
  Privacy (S\&P)}, pp.~3--18, 2017.

\bibitem{property-inference}
L.~Melis, C.~Song, E.~De~Cristofaro, and V.~Shmatikov, ``Exploiting unintended
  feature leakage in collaborative learning,'' in {\em Proc. IEEE Symp. Secur.
  Privacy (S\&P)}, pp.~691--706, 2019.

\bibitem{model-inversion}
M.~Fredrikson, S.~Jha, and T.~Ristenpart, ``Model inversion attacks that
  exploit confidence information and basic countermeasures,'' in {\em Proc. ACM
  SIGSAC Conf. Comput. Commun. Secur.}, pp.~1322--1333, 2015.

\bibitem{abadi2016deep}
M.~Abadi, A.~Chu, I.~Goodfellow, H.~B. McMahan, I.~Mironov, K.~Talwar, and
  L.~Zhang, ``Deep learning with differential privacy,'' in {\em Proc. ACM
  SIGSAC Conf. Comput. Commun. Secur.}, pp.~308--318, 2016.

\bibitem{dwork2014algorithmic}
C.~Dwork, A.~Roth, {\em et~al.}, ``The algorithmic foundations of differential
  privacy,'' {\em Foundations and Trends{\textregistered} in Theoretical
  Computer Science}, vol.~9, no.~3--4, pp.~211--407, 2014.

\bibitem{9069945}
K.~Wei, J.~Li, M.~Ding, C.~Ma, H.~H. Yang, F.~Farokhi, S.~Jin, T.~Q.~S. Quek,
  and H.~V.~Poor, ``Federated learning with differential privacy: Algorithms
  and performance analysis,'' {\em IEEE Trans. Inf. Forensics Secur.}, vol.~15,
  pp.~3454--3469, 2020.

\bibitem{9666041}
H.~H. Yang, Z.~Chen, T.~Q.~S. Quek, and H.~V. Poor, ``Revisiting analog
  over-the-air machine learning: The blessing and curse of interference,'' {\em
  IEEE J. Sel. Top. Signal Process.}, vol.~16, no.~3, pp.~406--419, 2022.

\bibitem{9895450}
N.~Zhang, M.~Tao, J.~Wang, and S.~Shao, ``Coded over-the-air computation for
  model aggregation in federated learning,'' {\em IEEE Commun. Lett.}, vol.~27,
  no.~1, pp.~160--164, 2023.

\bibitem{6573232}
M.~Goldenbaum and S.~Stanczak, ``Robust analog function computation via
  wireless multiple-access channels,'' {\em IEEE Trans. Commun.}, vol.~61,
  no.~9, pp.~3863--3877, 2013.

\bibitem{9835537}
Q.~Li, Y.~Diao, Q.~Chen, and B.~He, ``Federated learning on non-{IID} data
  silos: An experimental study,'' in {\em Proc. IEEE ICDE}, pp.~965--978, 2022.

\bibitem{9446488}
S.~Hu, X.~Chen, W.~Ni, E.~Hossain, and X.~Wang, ``Distributed machine learning
  for wireless communication networks: Techniques, architectures, and
  applications,'' {\em IEEE Commun. Surv. Tutorials}, vol.~23, no.~3,
  pp.~1458--1493, 2021.

\bibitem{qu2023learn}
Y.~Qu, X.~Yuan, M.~Ding, W.~Ni, T.~Rakotoarivelo, and D.~Smith, ``Learn to
  unlearn: A survey on machine unlearning,'' {\em \rm arXiv preprint
  arXiv:2305.07512}, 2023.

\end{thebibliography}

\vfill

\end{document}